# CH(A) Radical Formation in Coulomb Explosion from Butane Seeded Plasma Generated with Chirp-Controlled Ultrashort Laser Pulses

*Karoly Mogyorosi[1], Balint Toth[1], Krisztina Sarosi[1,2], Barnabas Gilicze[1], Janos Csontos[1], Tamas Somoskoi[1], Szabolcs Toth[1], Prabhash Prasannan Geetha[1], Laszlo Toth[1], Samuel S. Taylor[3], Nicholas Skoufis[3], Liam Barron[3], Kalman Varga[3], Cody Covington[4], Viktor Chikan[1,5,6]*

[1]ELI ALPS, ELI-HU Non-Profit Ltd, Wolfgang Sandner u. 3, H-6728 Szeged, Hungary

[2]Department of Optics and Quantum Electronics, University of Szeged, Dóm tér 9, H-6720 Szeged, Hungary

[3]Department of Physics and Astronomy, Vanderbilt University, Nashville, Tennessee, 37235, USA

[4]Department of Chemistry, Austin Peay State University, Clarksville, USA

[5]Department of Chemistry, Kansas State University, Manhattan, Kansas 66506-0401, United States

[6]ASML, 17082 Thornmint Ct, San Diego, CA 92, United States

ABSTRACT

We experimentally studied the formation of CH(A) radicals in butane seeded plasma generated with chirp-controlled ultrashort laser pulses (~760 μJ/pulse, 890 nm, 1 kHz, 8 fs). The focused beam with high peak intensity (~$10^{14}$–$10^{16}$ W/cm$^2$) caused Coulomb explosion (CE). The time dependent emission spectra were observed with the Fourier-transform Visible spectroscopy (FTVis) step-scan method. The average signal intensity decreased with the chirp in the Ar$^+$ > C$_2$ > H-α ~ CH(A) order with a plateau for CH(A) in the –200 to –100 fs$^2$ range. The short rise time of the CH(A) emission signal, the monoexponential emission decay and the nearly constant rotational and vibrational temperatures of the CH(A) radicals (~3000 K and ~3800 K) all support their formation as a primary product. Our TDDFT calculations predict that CH and many other fragments can be formed beyond CE at ~7×$10^{14}$ W/cm$^2$ intensity. The average charge of CH (+0.6) and its relative abundance (0.5%) support the formation of detectable CH(A) within 120 fs.

## Graphical abstract

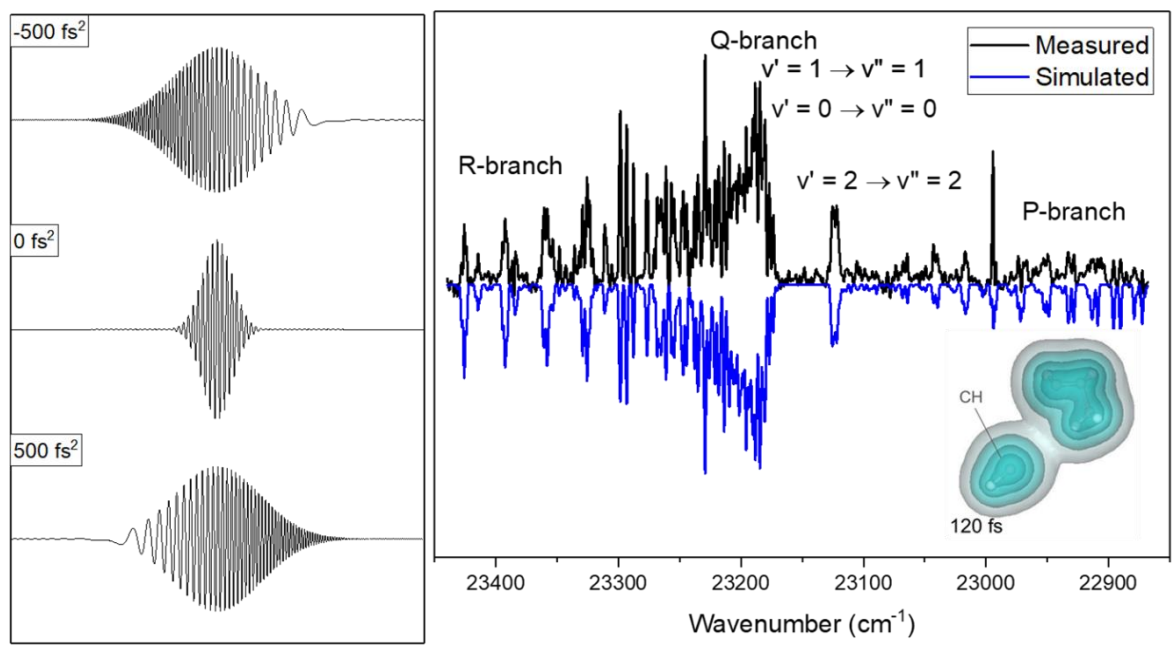

INTRODUCTION

The CH (methylidyne) molecule, which was among the first identified radicals,[1] plays an important role in chemical evolution in space under different astrophysical conditions.[2,3] CH radical formation is also significant in hydrocarbon flames[4–7] and plasma-enhanced chemical vapor deposition processes.[8] CH radicals can be also generated in chemical reactions induced by high energy photons, when the photodissociation of small hydrocarbon molecules occur. The blue light emission from the $A^2\Delta - X^2\Pi$ transition was extensively studied in many publications.[9–11] The experimental data were obtained from dissociative excitation of simple aliphatic hydrocarbons ($CH_4$, $C_2H_2$, $C_2H_4$ and $C_2H_6$).[10] The spectroscopy data allow a detailed characterization of the produced CH(A) radicals in plasma experiments.[2,12] Via the application of ultrashort femtosecond laser pulses the Coulomb explosion (CE) process produces a mixture of ionized and neutral fragments from hydrocarbons. The CE of smaller molecules primarily yields ionized species[13–17], studied by mass spectrometry based detection techniques, such as time-resolved Coulomb explosion imaging (CEI).[14] However, as the number of electrons and nuclei increases, it is reasonable to assume that the probability of the formation of neutral fragments will increase as shown in a few cases.[18] Initial experiments with $CH_4$ in an intense laser beam ($10^{14}$ W/cm$^2$) suggest that CH could be produced from neutral dissociation via a super-excited state (SES) of $CH_4$.[13] The SES can be generated via multiphoton excitation from a neutral molecule that will subsequently result in photodissociation. There are only few publications on experimental evidence for the formation of neutral fragments via CE – when single ionization is followed by photodissociation[18] – or the SES mechanism.[13]

Time-Dependent Density-Functional Theory (TDDFT) calculations were already used for the simulation of the ionization and Coulomb explosion of small organic molecules in the presence of strong laser pulses.[19–23] The spectra of ejected protons from hydrocarbons (methane and 1,3-butadiene) calculated by the TDDFT approach was in good agreement with the experimental data.[23] The pulse length and alignment-dependent ionization of ethylene and acetylene were also well described with this method.[21,24] However, neutral species from Coulomb explosion may still appear, which is an aim of this work experimentally and theoretically to prove with butane as a precursor molecule.

MATERIALS

High purity argon (Messer, 5.0, 99.999%, $H_2O$ <3 ppm (V/V), $O_2$ <2 ppm (V/V), $N_2$ <5 ppm (V/V), $\Sigma C_nH_m$ <0.2 ppm (V/V), $CO_2$ <0.5 ppm (V/V), 3 bar) and n-butane/Ar mixture (Messer, $C_4H_{10}$, 3 v/v%, ±2 rel %, 3 bar) are used in these experiments. For the rise time measurements, n-butane/He mixture (3 v/v% n-butane content, Messer) and methane/He mixture (3 v/v% methane content, Messer) gas mixtures were also applied.

EXPERIMENTAL SETUP

The molecules are introduced into a 12" hemispherical vacuum chamber (SP1200S-316LN-EP, Kurt J. Lesker) that is evacuated with a turbo pump (STP-2207C, Edwards). The piezo valve (Amsterdam Cantilever Piezo Valve, ACPV3 and the Electronic Driver Unit EDU5, operating at 0–200 V opening voltage and 2–400 µs opening time) allows the injection of the molecules into the reaction chamber from the top of the chamber. The piezo valve opens at 1 kHz frequency (used at 200 V and 40 µs), and is synchronized with the SYLOS2 laser at ELI ALPS (890 nm central wavelength, 8 fs pulse duration, ~0.76 mJ pulse energy, 1 kHz repetition rate). The pressure is in the $3.0–3.5\times10^{-3}$ mbar range when the piezo valve is open in the pulsed mode ($10^{-7}–10^{-8}$ mbar when closed).

The experiments using the SYLOS2 laser source are carried out at the ELI-ALPS, which provides >30 mJ, <8 fs pulses at 1 kHz repetition rate. The system was recently improved to provide Gaussian beam shape with a waist diameter of 34 mm at FWHM. The output of this laser system is sampled by a wedge pair (Figure 1a) to provide the necessary power for the photodissociation experiments. The reflection from the first uncoated surface provided ~760 µJ pulse energy for the experiments. An imaging telescope is used to reduce the beam size to ~10 mm which was suitable for the experiment. The focal plane of the telescope is located at the middle of a 1 m long vacuum tube in order to avoid nonlinear effects in air. A fused silica wedge pair and 8 bounces on a chirp mirror pair (+100 $fs^2$ GDD in each reflection) were used to fully compress the pulses before the interaction chamber (Figure 1b).

(a)

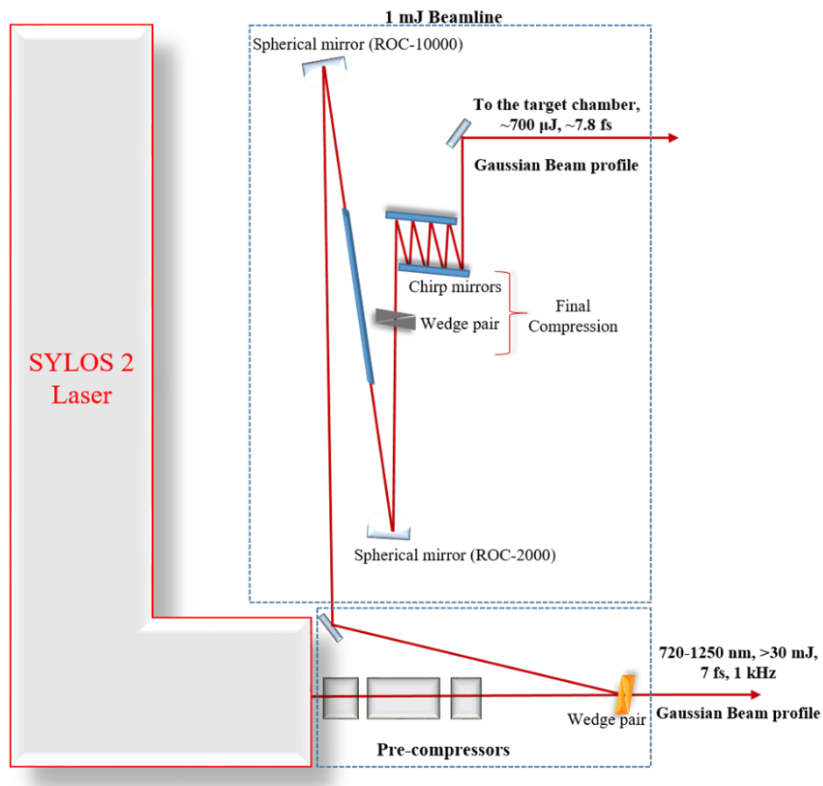

(b)

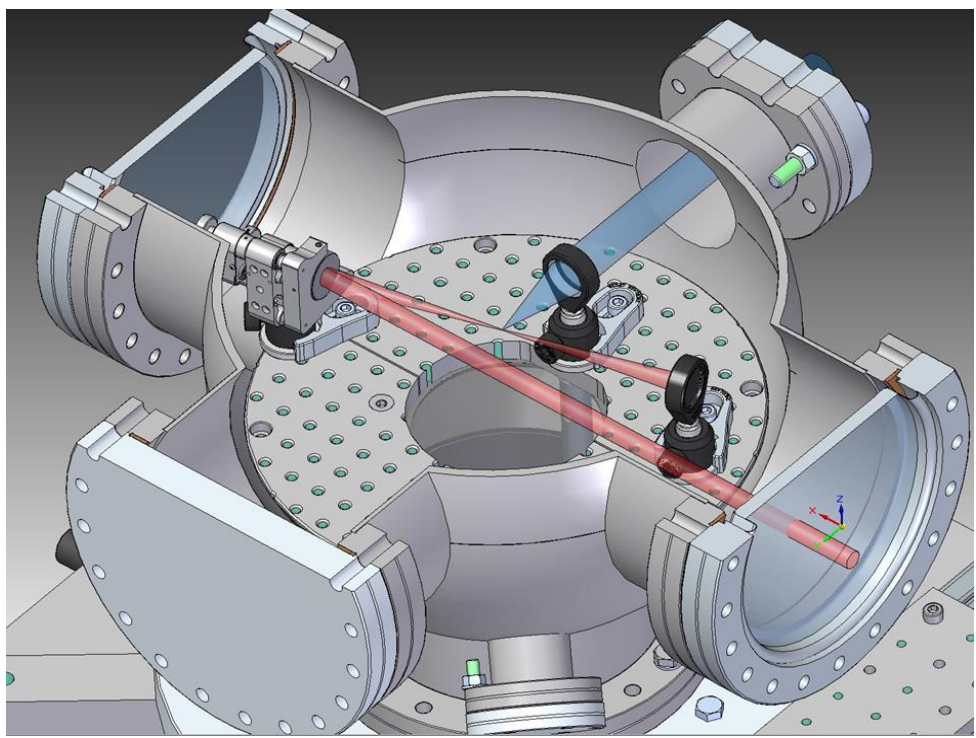

**Figure 1.** (a) Schematic view of the SYLOS2 laser system and 1 mJ beam line for the photodissociation experiments, (b) the reaction chamber used in the photodissociation experiments (red beam represents the SYLOS2 beam and the blue one the collected fluorescent light)

In these experiments, the Gas Phase Reaction Control (GPRC) vacuum chamber was used for the characterization of photofragments from the CE of small organic compounds (Figure 1b). In this method, molecules from pre-mixed gases are introduced via a pulsed piezo valve synchronized with the laser.

The trigger signal timing for the piezo valve is provided with a delay generator (DG645, Stanford Research Systems) that allows the maximum overlap between the appearance of the molecules and the ultrashort laser pulses. The height of the beam is set with a periscope and introduce the laser pulses through a small, fused silica window (2 mm thickness, 25 mm clear aperture) into the vacuum chamber. The beam is focused with a spherical mirror under the piezo valve (CM254-100-P01, Concave mirror, f = 100.0 mm, Protected silver). The focal spot is positioned with the fine adjustment of the mirror in a picomotor controlled mirror mount (UHV Picomotor Piezo Clear Edge Mount, 1.0 in., 8821-L-UHV and 8742-4-KIT, NewFocus).

The emitted light is collected from the plasma with a collimating lens placed perpendicularly to the direction of beam propagation (Figure 1b). There are two spectrometers in the experimental setup to cover a variety of experimental needs. The overview spectra are measured with the fiber coupled Ocean Optics QEPro spectrometer (200–1100 nm wavelength range, time averaged) or with the Bruker Vertex80 FT-Vis spectrometer. The latter is used in step-scan mode with 5 ns or 2.5 ns time resolution in the 5–500 ns range and 1 cm$^{-1}$ spectral resolution (further details in our previous publications).[12,25] In order to maximize the CH(A) specific detection at about 431 nm, we use shortpass and band-pass filters (FES0450 and FB430-10, Thorlabs) in the Bruker Vertex80 spectrometer. These CH(A) fluorescence measurements were carried out with 100 co-additions in step-scan mode and the measurement required about 45 minutes. A CaF$_2$ (UV-NIR, 50 000 – 4000 cm$^{-1}$) beam splitter is used in this spectrometer and a Hamamatsu PMT (H10721-113) with a fast pre-amplifier (Ortec 9306, rise time 300 ps) for the sensitive detection of the fluorescent light from the plasma. The Instrument Response Function (IRF) and the rise time measurements were also performed with the HR1 alignment laser (1030 nm, 7 fs, 1 kHz, 1 mJ/pulse), applying a Tektronix oscilloscope (MSO 71254C, 12.5 GHz).

The analysis of the time dependent spectra were processed in PYTHON. The simulation of the Ar$^+$ emission spectrum based on 3D modelling of the ultrashort laser pulses and calculation of the ionization rate was performed in MATLAB.

The broad spectrum of the SYLOS2 laser in the 750–1100 nm wavelength range and the temporal shape of the pulses are shown in Figure S1.[26] The pulse duration at the interaction point was characterized and optimized by the chirp scan technique, which provided a 7.9 fs compressed pulse shape (Figure S1b).[27,28]

Pulse compression, chirp correction and pulse shaping were manipulated by a Dazzler or Acousto-Optic Programmable Dispersive Filter (AOPDF). This accurate pulse manipulation was accomplished with the longitudinal interaction between polychromatic acoustic and optical waves in a birefringent crystal. In the chirp scan measurements, the second harmonic of the fundamental wavelength is produced with a beta barium borate (BBO) crystal and the second harmonic spectra was scanned over a GDD range. Using an iterative algorithm, we can retrieve the fundamental spectrum, spectral phase and pulse duration from this measured SHG spectra.

The pulse duration determination at different GDD values are shown in Figure S2 and the retrieved output spectrum and spectral phase of SYLOS2 (Figure S3). The temporal distribution of the chirped pulses are calculated from the retrieved complex spectrum (Figure S2). As we can see in Figure S3, the spectral amplitude has a regular shape, and the spectral phase is quasi-constant in the entire wavelength range (700 –1100 nm). For this reason, we are very close to the transform limit in case of GDD 0 $fs^2$ value.

We measured the beam diameter and shape in the focal plane outside the vacuum chamber through the same type of fused silica window that is in the vacuum chamber. The beam diameter was found to be larger (55 μm) than the ideal size and not fully Gaussian in shape. Considering this beam diameter and shape in the focal plane, the calculated peak intensity is $1.2 \times 10^{16}$ W/cm$^2$.

TDDFT simulations

The electron dynamics in the simulations were modeled using TDDFT[29] on a real-space grid with real-time propagation. Core electrons, which are difficult to handle computationally, were represented using norm-conserving Troullier-Martins pseudopotentials.[30] At the beginning of the TDDFT calculations, the ground state of the system was prepared by performing a density functional theory calculation. Next, the time-dependent Kohn-Sham orbitals were determined by solving the time-dependent Kohn-Sham equation. All orbitals were represented on a real-space grid inside a large simulation volume (29.7 × 29.7 × 29.7 Å) with a grid spacing of 0.3 Å. Each Kohn-Sham orbital was time-propagated using the evolution operator in the form of the fourth-order Taylor expansion[31] with a small time step of 1 attosecond until the end of the 120 fs simulation window. The TDDFT simulations are carried out using the adiabatic local density approximation (ALDA) with the parametrization of Perdew and Zunger[32]. The motion of the ions in the simulations was treated classically using quantum forces in the framework of the Ehrenfest dynamics. Complex absorbing potentials (CAP)[20] were used at the boundaries to prevent the reflection of the wave function from the boundaries. We have developed several different computational approaches[19,33,34] to make our TDDFT program more efficient and accurate, and in previous work have shown this approach to be successful in describing the Coulomb explosion of molecules.[19–23]

PGOPHER spectroscopy software was used for calculating the vibrational and rotational temperatures from the emission spectra from CH(A) radicals in the 22,500-24,000 cm$^{-1}$ spectral range.[35] We normalized the average high resolution spectra to the strong argon (Ar$^+$) emission line at 22,995 cm$^{-1}$ (434.9 nm) and calculated the argon background corrected spectra for each data point. We averaged the spectra in 50 ns time windows, such as 50–100 ns, 100–150 ns, 150–200 ns, 200–250 ns and 250–300 ns, and these average spectra were applied for the simulations. We collected the molecular constants from the literature and generated the line list from a reference CH(A) spectrum.[36,37] We set the initial vibrational and rotational temperatures to be 2500 K. First, we optimized the scale, the baseline and offset parameters and subsequently the Gaussian parameter, and finally determined the rotational and vibrational temperatures.

RESULTS AND DISCUSSION

In the first experiments, we explored the influence of the distance of the focal spot and the piezo valve slit by tilting the focusing mirror. In Figure 2a, the emission spectrum of the plasma indicates the presence of two diatomic radicals, i.e. CH(A) at 431 nm (CH($A^2\Delta \rightarrow X^2\Pi$), CH(A-X)) and C$_2$ at 450–475 nm, 500–520 nm and 550–570 nm (D$^3\Pi$-A$^3\Pi$, Swan bands).[38,39] We also identified two atomic lines of hydrogen (H-α, 656 nm and H-β, 486 nm). The emission spectrum of Ar$^+$ shows an intense peak at 434.9 nm (22,995 cm$^{-1}$), which overlaps with the investigated emission band of CH(A) (22,500–24,000 cm$^{-1}$, 417–444 nm, Figure 2a). We took images of the plasma by changing the distance between the piezo valve and the focal spot by moving the motorized mirror mount from the 1.80 mm to the 2.60 mm position (Figure S4). At the optimum chirp value (shortest pulse duration, ~8 fs), plasma generation is highly efficient and the emitted light from argon plasma can be detected with the QEPro spectrometer (Figure

2a). We monitored the main peak intensity of argon at about 435 nm in the overview spectrum and set the most suitable delay between the valve opening starting time and the pulse arrivals. Tilting the spherical mirror mount allowed us to systematically change the distance between the focal spot and the valve slit. In the case of butane, we monitored the emission spectrum from 2.0 mm to 2.4 mm distance and observed a gradual increase in the peak intensity at 431 nm. However, this also resulted in higher intensities in the argon emission line (Figure S5). We selected the 2.2 mm distance at which the CH(A-X) emission intensity is near to the maximum and the contribution from argon is not too high (Figure S5).

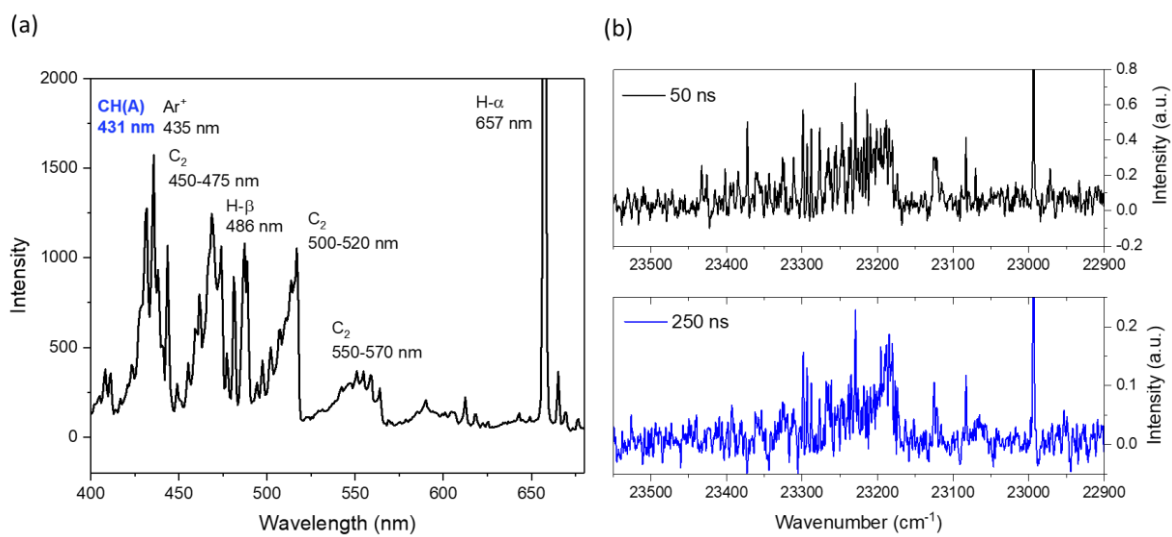

**Figure 2.** (a) Overview emission spectrum measured from the brightest butane-argon plasma with the CH(A), $C_2$ emission bands and H atomic lines (b) High resolution emission spectra from argon-butane plasma measured at 50 and 250 ns time points (1 cm$^{-1}$ spectral resolution)

The plasma emission spectrum from the argon-butane mixture was measured with high spectral resolution (1 cm$^{-1}$) in the 22,000–25,000 cm$^{-1}$ spectral range in the step-scan mode (Figure 2b). The applied narrow band filter is suitable for measurements in the 430 ± 5 nm (23,530–22,990 cm$^{-1}$) spectral range. Although this is a narrow observation window, one intense emission line from Ar$^+$ ions at 435 nm (434.9 nm, 22,995 cm$^{-1}$) was also measured along with three lower intensity argon lines (23,083 cm$^{-1}$, 23,373 cm$^{-1}$ and 23,433 cm$^{-1}$). In the first 10 ns, the background from argon emission is significant compared to the typical CH(A-X) emission (see also the instrument response measurement with the second harmonic pulses of the SYLOS beam at 445 nm).

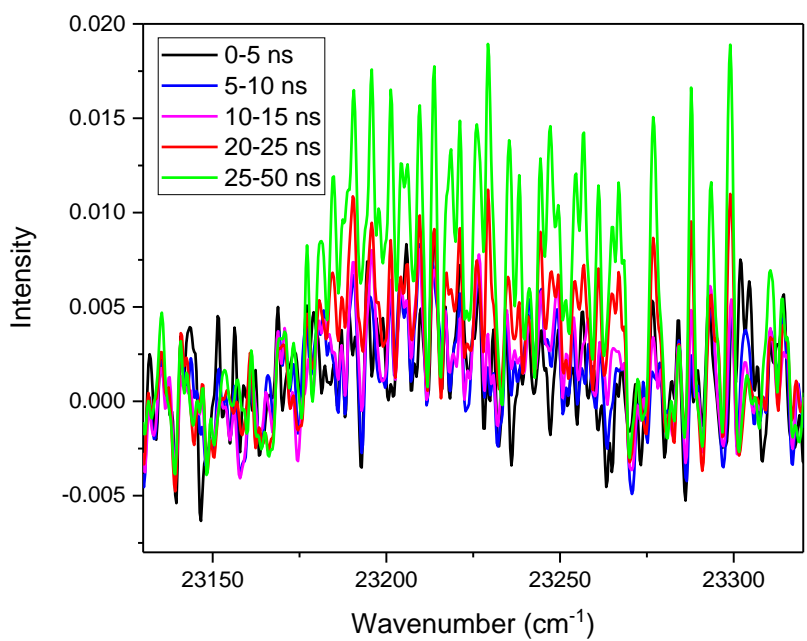

(b)

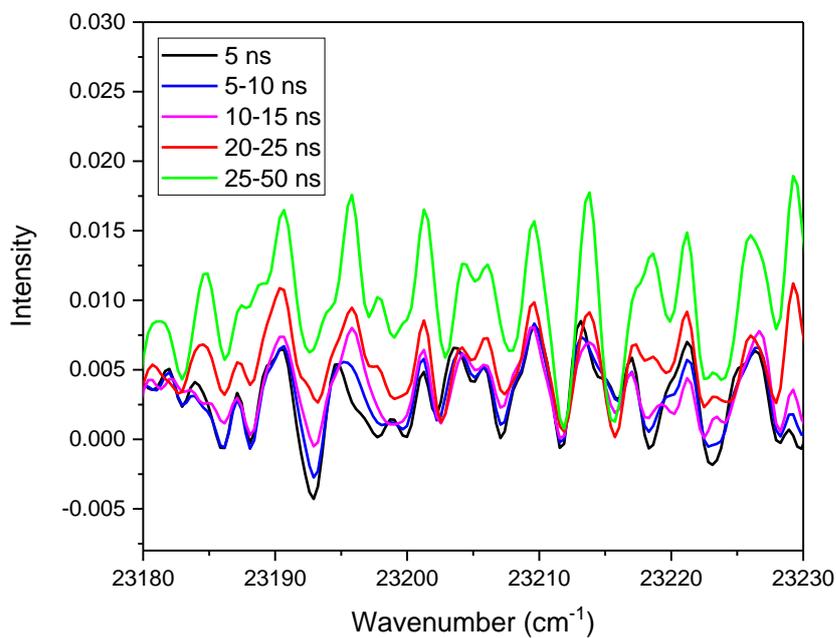

**Figure 3.** The average CH(A) emission spectrum at 0–25 ns and the early spectra in (a) 23,130-23,320 cm$^{-1}$ and (b) 23,180–23,230 cm$^{-1}$ spectral range (background corrected spectra)

One could first recognize the CH(A-X) spectrum shape at about 5–10 ns as it evolves from the significant argon emission background. The CH(A) average spectrum and the argon background spectrum were calculated for the investigated time window (ie 25-50 ns) from parallel measurements. The average argon spectrum was then normalized to the sample spectrum at 22,995 cm$^{-1}$ (434.9 nm). During background correction, the normalized argon spectrum was subtracted from the spectrum of the sample. In Figure 3, the background corrected average CH(A) spectrum (25-50 ns) is compared with the early spectra measured at GDD 0 fs$^2$ value. We can conclude from the observed spectra that the presence of CH(A) radicals is likely from the 5 ns time point. We performed additional measurements with methane-helium mixture and 2.5 ns time resolution in order to confirm the very early presence of the CH(A) radicals from CE. The spectra showed their presence from the first detection point at 2.5 ns (Figure S6).

The group delay dispersion (GDD) scan was applied with the Dazzler in the −1000 to +1000 fs$^2$ GDD range. At a given GDD value, we measured the average overview emission spectra with the QEPro spectrometer (Figure S7). The spectra were analyzed for the Ar$^+$ (435 nm), CH(A) (431 nm), C$_2$ (515 and 550 nm) and H-alpha (656 nm) emission intensities from butane-argon plasma and the Ar$^+$ (435 nm) peak also from pure argon plasma (Figure 4). As expected, the highest emission intensity was observed at the shortest transform limited pulses where the GDD value was 0 fs$^2$. For better comparison, all these spectrum intensity values were normalized for their maxima. It can be seen from Figure 4a that all intensities decrease rapidly in the ±60 fs$^2$ GDD range. The sharpest decrease was observed for Ar$^+$ emission, when the GDD was lower or higher than the optimum. In pure Ar, the GDD dependent intensity profile of Ar$^+$ emission was more symmetric compared to that obtained for Ar$^+$ emission in the presence of butane. However, this was most likely caused by some contribution of other components emitting light at 435 nm, such as CH(A) radicals. Considering the laser beam diameter (d = 10 mm), the focal length of the spherical mirror (f = 100 mm), the measured focal spot diameter 55 μm, and the pulse energy (0.76 mJ/pulse), the calculated peak intensity is 1.2×10$^{16}$ W/cm$^2$. This very high peak intensity results in CE in the sample as it requires at least 10$^{14}$ W/cm$^2$ peak intensity.[18,21,23,24] Liu and co-workers studied the dissociative ionization and CE of bromocyclopropane in an intense femtosecond laser field (800 nm, 1 kHz repetition rate, 50 fs, focusing with a biconvex lens, f = 40 cm; up to 2×10$^{14}$ W/cm$^2$).[18] The C-C and C-Br bonds were cleaved and the main ionized products (H$^+$, C$^{2+}$, CH$_m^+$, C$_2$H$_m^+$, C$_3$H$_3^+$, C$_3$H$_5^+$, Br$^+$, CH$_2$Br$^+$, m = 0-3) were detected with the TOF-MS technique.

The ionization energy of argon is 15.8 eV, which is very high compared to the energy of a single photon of the laser beam (at 890 nm central wavelength, 1.4 eV). The focused beam with high peak intensity caused the ionization of the argon atoms, and carbon atoms were also ionized (emission from C$^+$ ions at 426.8 nm, 23,429 cm$^{-1}$ at 0 fs$^2$ GDD). The Ar$^+$ emission intensity rapidly decreases when the GDD value is at ±60 fs$^2$ with an almost 40% intensity change. At +200 fs$^2$ GDD value, the pulse duration is about 100 fs, which results in 1.0×10$^{15}$ W/cm$^2$ peak intensity (Figure S8). At −1000 fs$^2$ GDD value, the calculated pulse length is about 470 fs, which provides 1.8×10$^{14}$ W/cm$^2$ peak intensity.

We simulated the Ar$^+$ emission spectrum based on 3D modelling of the ultrashort laser pulses and calculation of the ionization rate (as a function of time and space) in the focal plane using the Ammosove–Delone–Krainov model.[40,41] The simulation of the pulses in the spectral time domain is based on experimental values of the spectral amplitude and spectral phase measured by the chirp-scan method. Regarding the spatial distribution, an ideal Gaussian beam is assumed. For a given pulse shape and peak intensity, we calculate the accumulated ionization fraction of the medium and integrate it in space. This ionized volume must be proportional with the intensity of the Ar$^+$ in the emission spectra. The simulated normalized ionization volumes as a function of GDD are shown in Figure 4b for different peak intensity values. The simulation result at 3.0×10$^{16}$ W/cm$^2$ is in good agreement with the measured Ar$^+$ emission profile in the −200 to +200 fs$^2$ GDD range. Therefore the determined maximum peak intensity from the simulation is close to the measured value.

Surprisingly, the production of the detected other spectral components are less sensitive for the group delay dispersion change. The sensitivity order is Ar$^+$ > C$_2$ > H-α ~ CH(A). As a butane molecule (C$_4$H$_{10}$) contains four carbon atoms, the C$_2$ radicals in the excited state can be produced in different ways after the C-H and C-C bonds are broken, if a single C-C bond remains from the parent molecule. The hydrogen atoms are ejected with high probability, which results in very high H-α emission intensities (see Figure S7). It is also notable that the GDD dependent intensity profiles of CH(A) and H-alpha are asymmetric with higher values in the negative chirp range with a shoulder between -200 and -100 fs$^2$ GDD values. This means that both the pulse duration and the sign of chirp are important. The negatively chirped pulse means that the instantaneous frequency decreases with time (or its wavelength increases). Thus, the molecules are affected first by the blue part (higher energy photons) and later the red part of the pulse spectrum. This could make different bond breaking mechanisms more efficient in producing CH(A) excited radicals and H atoms. The increased CH(A) emission for negative GDD could be also explained by enhanced stimulated emission, which is favored by negative GDD.[42]

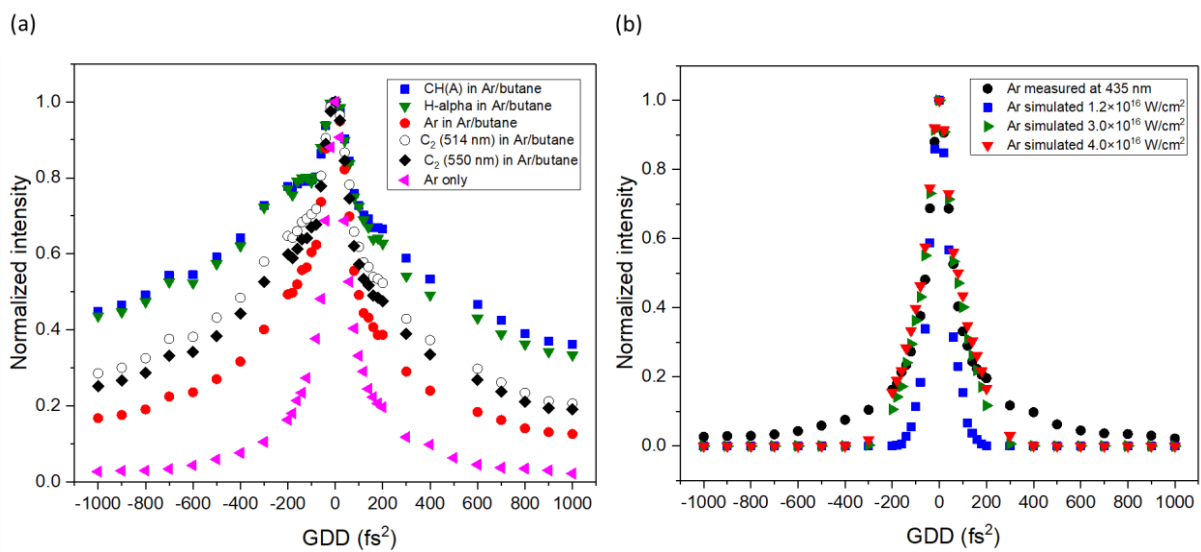

**Figure 4.** (a) The measured GDD dependent normalized intensities in the CH(A) (431 nm), $C_2$ bands (515 and 550 nm) and $Ar^+$ (435 nm), H-alpha (656 nm) lines in the overview spectra from butane-argon sample and the $Ar^+$ (435 nm) only in pure argon and (b) the simulated argon spectra at three maximum peak intensity values at GDD 0 $fs^2$ ($1.2\times10^{16}$ W/cm$^2$, $3\times10^{16}$ W/cm$^2$ or $4\times10^{16}$ W/cm$^2$)

The emission spectra measured at –1000, –200, 0, 200 and 1000 $fs^2$ GDD are shown in Figure S7. The CH(A) spectrum is less contaminated with the $Ar^+$ emission at the lower and higher GDD values. We were able to characterize the kinetics of the signal evolution for the H-α line at 656 nm (Figure S9). The signal rapidly reaches its maximum within 5 to 10 ns, and decays within 20 ns. This H-α signal time profile is close to the expected Instrument Response Function (IRF, determined by measuring the SHG pulse from the SYLOS2 laser, FWHM is about 9.5 ns, Figure S9), which is plausible considering the short lifetime of H-α emission (17 ns)[43]. Similarly, the $Ar^+$ emission increases in 10 ns, since its detection is limited by the slower instrument response compared to the very rapid $Ar^+$ ion formation in the first few hundred picosecond time range.[44]

The argon background corrected spectra were also calculated by adjusting the $Ar^+$ line intensity at 22,995 cm$^{-1}$ in pure argon to the $Ar^+$ line intensity measured for the given time point in butane-argon mixture. The butane-argon spectrum was then corrected with this adjusted argon background spectrum (Figure 5). One can recognize the P-, Q- and R-branches of the CH(A) emission in the measured spectrum, although the P-branch is not fully covered applying the FB430-10 bandpass filter.

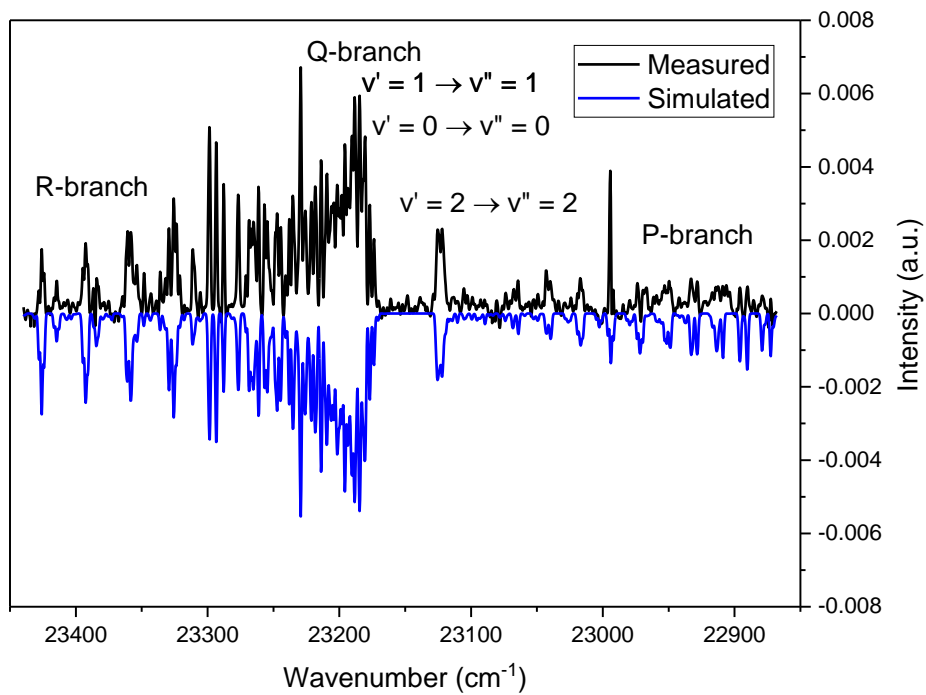

**Figure 5.** Measured argon background corrected and simulated emission spectra of CH(A) from the butane-argon sample at +200 fs$^2$ GDD in the 150–200 ns time range

The background corrected spectra were also used for the calculation of the kinetic traces at the −200 fs$^2$, 0 fs$^2$, 200 fs$^2$ and 500 fs$^2$ GDD (Figure 6). The integrated CH(A) signals (23,175-23,310 cm$^{-1}$) were already at 87%, 60%, 98% and 70 % of the maximum CH(A) intensity at 5 ns for the −200 fs$^2$, 0 fs$^2$, 200 fs$^2$ and 500 fs$^2$ GDD values, respectively (Figure S10). The rising period was in the 0–25 ns time interval for the 0 fs$^2$ GDD case. At −200 fs$^2$ and 200 fs$^2$ GDD, the CH(A) rise was only 10 ns and 5 ns, respectively. The decay curves were fitted with monoexponential functions with $t_1$ = 155 ± 4 ns, 98.0 ± 4.9 ns, 141 ± 6 ns and 168 ± 7 ns for the −200 fs$^2$, 0 fs$^2$, 200 fs$^2$ and 500 fs$^2$ GDD, respectively. The formation of the CH(A) signal is very rapid and has a similar time profile as the IRF. In order to further explore the evolution rate of the CH(A) radicals, we characterized the IRF and the signal measured directly with a fast PMT (Hamamatsu, H10721-20, 0.5 ns rise time), an Ortec pre-amplifier (9306, 1 GHz, 300 ps rise time) and a Tektronix oscilloscope (MSO 71254C, 12.5 GHz). The IRF of this detection scheme was measured directly by applying the ultrashort Second Harmonic (SH) pulses from the HR1 alignment laser (FWHM is 2.08 ± 0.02 ns from a Gauss fit, rise time 1.0 ns, Figure S11). We used butane-argon mixture for that purpose both in pulsed mode and also in static mode. The rise time of the signal in the selected narrow spectral range were 1.36 ns for Ar and 1.60 ns for the butane-argon sample in pulsed measurement mode (Figure S12a). These values were rather similar for the Ar (1.28 ns) and butane-argon sample (1.32 ns) in static measurement mode (Figure S12b). We can observe that the formation of the CH(A) radicals is obvious in this short time scale considering the broader signals and the remaining slower decaying component. Since the $Ar^+$ emission gives a significant contribution to the signal of the CH(A) emission, we also measured the signal evolution from He and butane-helium samples in static mode (Figure S13). In this case, the rise time was 1.68 ns for He and 1.36 ns for butane-helium samples. The butane-helium peak is also broader compared to the He peak and the remaining slower decaying CH(A) emission signal is more pronounced compared to the Ar and butane-argon comparison. It is possible that the formation of CH(A) radicals is within 100 fs after the pulse arrives, thus they are produced directly in the CE process. Long-lived electrons from the argon plasma could also cause the formation of CH neutral radicals in the first few hundred-picosecond time range due to collisions with the $CH_2^+$ ions via dissociative recombination.[44–46] The pressure is $3\times10^{-3}$ mbar in the vacuum chamber, when the piezo valve operates in the pulsed mode. Based on recent pressure profile measurements with the same type of piezo valve (ACPV3, 9 bar argon and 60 μs opening time)[47] and literature data, the estimated pressure is two magnitude lower compared to the input pressure (3 bar), a few 10 mbar at the focal spot.

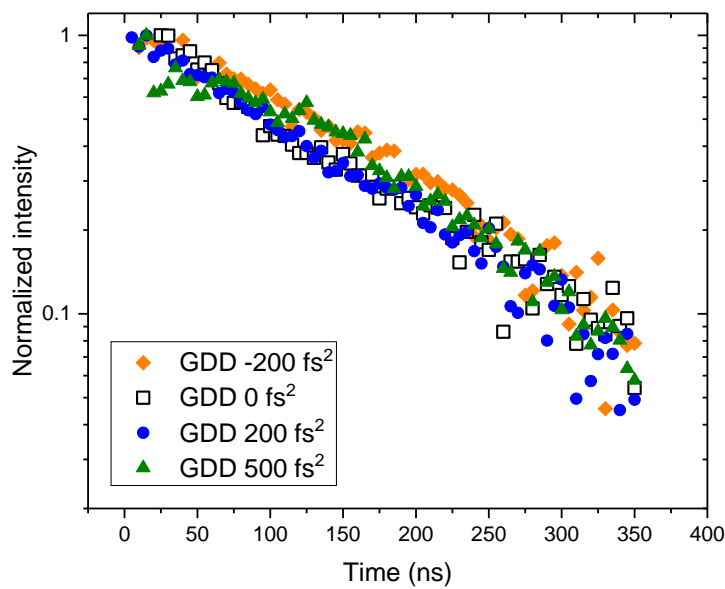

**Figure 6.** The GDD dependent exponential decay curves of the CH(A) emission integrated in the 23,175–23,310 cm$^{-1}$ spectral range with argon correction

Considering static gas conditions for Ar, 30 mbar pressure and 3500 K temperature, the number of collisions would be 1 in about 19 ns (see the Supporting information for the calculation details).[48] However, in our experiments the gas mixture is not in static condition due to the rapid gas expansion from the jet. The Coulomb explosion could also result in more frequent collisions, although the low concentration of the butane molecules (3 v/v%) increases the collision time for butane derived fragments. We can estimate the collision time of the ions in the plasma, for example for the $CH_2^+/CH_2^+$ ion-ion collisions to be 2.8 ns (eq 6 in the Supporting Information).[49] The formation of the CH(A) radicals from the collisions of the carbon containing fragments in the first nanosecond time after the CE is less likely compared to the direct CH(A) formation processes.

We observed a pronounced broader peak of the CH(A) emission kinetics only in the methane-helium measurement with a maximum at 60 ns (Figure S6 and S14). The formation of CH(A) radicals can be attributed to secondary reactions in the first few ten nanoseconds, caused by collisions between the nascent carbon and hydrogen atoms for example. In Figure 6, the decay curves show a very similar time profile (linear in semi-log plot). We assume that after the formation of the CH(A) radicals, their decay is mostly determined by their natural lifetime due to the low probability of chemical reactions at lower pressure in the vacuum chamber. The

155 ns time is still shorter than the natural lifetime of the CH(A) emission, which is reported to be in the 283–560 ns range.[50–52] The shorter lifetime could be the result of the fluorescence quenching due to the collisions with the argon atoms.[53] However, this might be also partly caused by the fact that the molecules from the jet are in rapid motion from the piezo valve towards the turbo pump. This displacement also causes shorter observed lifetimes.

PGOPHER spectroscopy software was used for calculating the vibrational and rotational temperature time profiles from the emission spectra from CH(A) radicals for the evaluation of temperature changes due to potential collisions (Figure 5).[35] [36,37] The average spectrum was calculated from parallel measurements for both the argon only and argon-butane mixture at a given GDD value for a selected time interval. It was necessary to average ten spectra to have a good quality spectrum for the fitting. Figure 5 shows an average measured spectrum for the 150–200 ns time period for the 200 $fs^2$ GDD value. The simulated spectrum is in good agreement with the obtained spectrum. Further fitting details are in our earlier publication.[54]

At the shortest pulse duration (~8 fs, 0 $fs^2$ GDD), we observed the emission line from $C^+$ ions (CII, 426.8 nm, Figure S15a). In some instances, the CH(A) emission was clearly recognizable in the 25–50 ns time window, for example at 0 $fs^2$ and +200 $fs^2$ GDD, however not observable at +500 $fs^2$ GDD (Figure S15a-c). The ionization of carbon atoms first occurred only under these conditions and was less likely at longer pulse durations (first ionization energy of carbon is 11.3 eV). The average spectrum from these early time points were still too noisy for fitting. Therefore, we could characterize the vibrational and rotational temperatures only from 75 ns (50–100 ns) up to 275 ns (250–300 ns time window).

We can see from Figure 7 that the vibrational temperatures are always higher at the earliest time point than the rotational temperatures. The vibrational temperatures are 3770 ± 80 K, 3850 ± 120 K, 3805 ± 75 K and 3810 ± 85 K at 75 ns and the rotational temperatures are 2965 ± 35 K, 3135 ± 55 K, 2930 ± 35 K and 2870 ± 35 K, respectively. The differences are small but apparently, there is a maximum at 0 $fs^2$ GDD for both temperatures. The rotational temperatures slowly decreased up to 300 ns, typically from ~3000–3100 K to 2000–2400 K. We simulated the average spectra with the lowest final rotational temperature for the 500 $fs^2$ GDD. However, the vibrational temperatures were practically all constant in this time range (50–300 ns). We noticed some increase in the vibrational temperature only for the sample with the shortest pulse duration from 3850 K to 4885 K (at 0 $fs^2$ GDD).

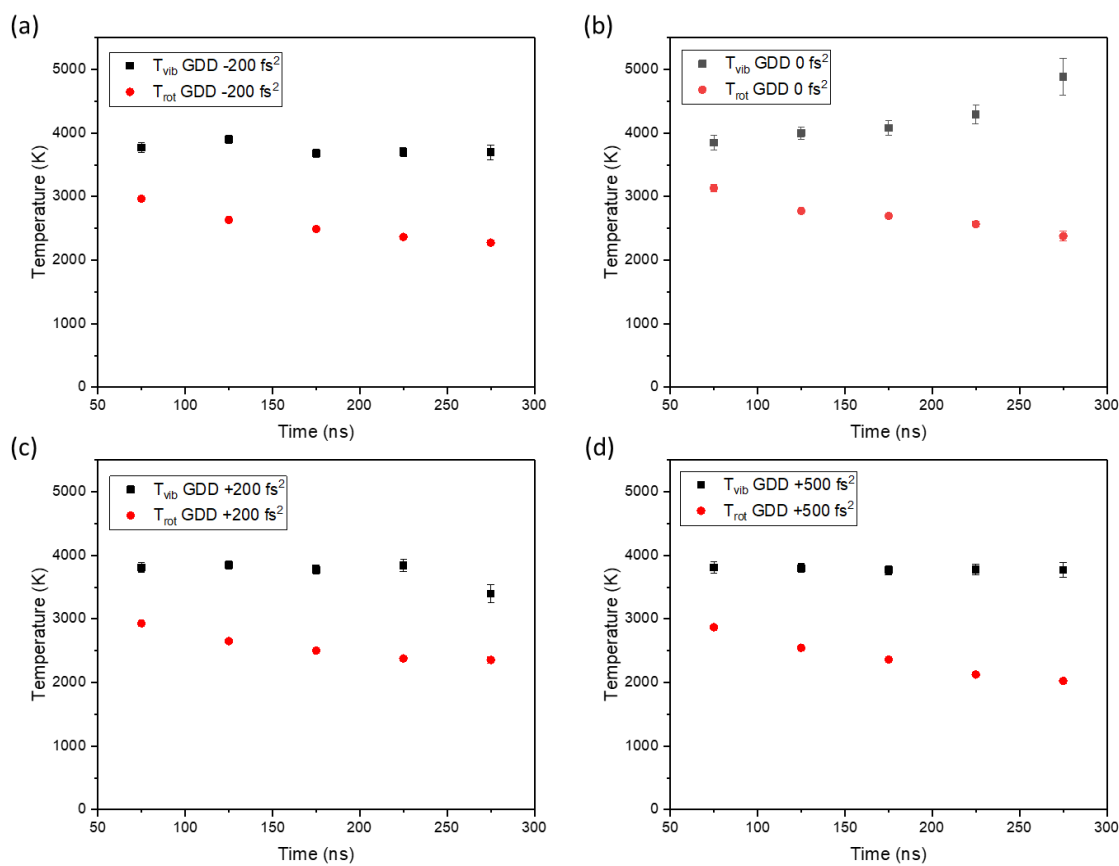

**Figure 7.** Vibrational and rotational temperature changes in the 50–300 ns time range at (a) -200 fs$^2$, (b) 0 fs$^2$, (c) +200 fs$^2$ and (d) +500 fs$^2$ GDD

Lindner and co-workers determined lower $T_{rot}$ = 1450 K (193 nm excitation) and $T_{rot}$ = 2090 K (248 nm excitation) temperatures for the multiphoton dissociation of bromoform.[55] They measured $T_{vib}$ = 2400 K temperature at lower laser intensity. However, at the highest applied pulse energy (120 mJ), the $T_{rot}$ = 1850 K and $T_{vib}$ = 3190 K temperatures were determined with 248 nm excitation from bromoform introduced in a supersonic jet. We assume that there are still collisions between the generated atoms and ions in the first few hundred nanosecond time range. Since the rotational energy transfer is much faster, this is likely caused by the collisions of the CH(A) radicals with argon atoms.[56] However, collisions with nascent carbon and hydrogen atoms may result in the formation of CH(A) radicals at higher vibrational temperatures in secondary condensation reactions after 100–150 ns, especially when the shortest pulse duration results in reactive nascent atoms in high concentration at the GDD 0 fs$^2$.

We determined experimentally that the ultrashort laser pulses (7.9 fs) with the measured pulse energy (0.76 mJ) represent very high electric field strength in the focal spot (~40 V/Å) and the peak intensity is about ~$2\times10^{16}$ W/cm$^2$ based on the Ar$^+$ emission simulations. The TDDFT simulations were performed to compare the experimentally observed emission spectra in the first time point of detection (<10 ns) with the theoretically predicted fragment distribution. At the beginning of the time dependent calculations, the velocity of the ions were initialized according to the Boltzmann distribution corresponding to 300 K. This random velocity initialization allows for different fragmentations during the Coulomb explosion. If the ions' velocity would be frozen to a given value initially, the calculations would produce exactly the same fragmentation. The other factors that control fragmentation is the orientation of the molecule and the parameters (intensity, shape, frequency) of the laser pulse (see Figure S16). In the calculations, the laser parameters are chosen to match the experiments and the laser is polarized in the *x* direction.

The simulations were first performed at a higher electric field strength, e.g. 40 V/Å, which resulted in an almost complete dissociation of the bonds in butane with 7.8 fs (full width at half maximum) ultrashort pulses. However, at 7.26 V/Å electric field strength ($7\times10^{14}$ W/cm$^2$, Figure S16) we started to observe the formation of different fragments from CH to $C_4H_6$ up to the end of the simulation time window (120 fs, Figure 8 and Coulomb explosion animation 1).

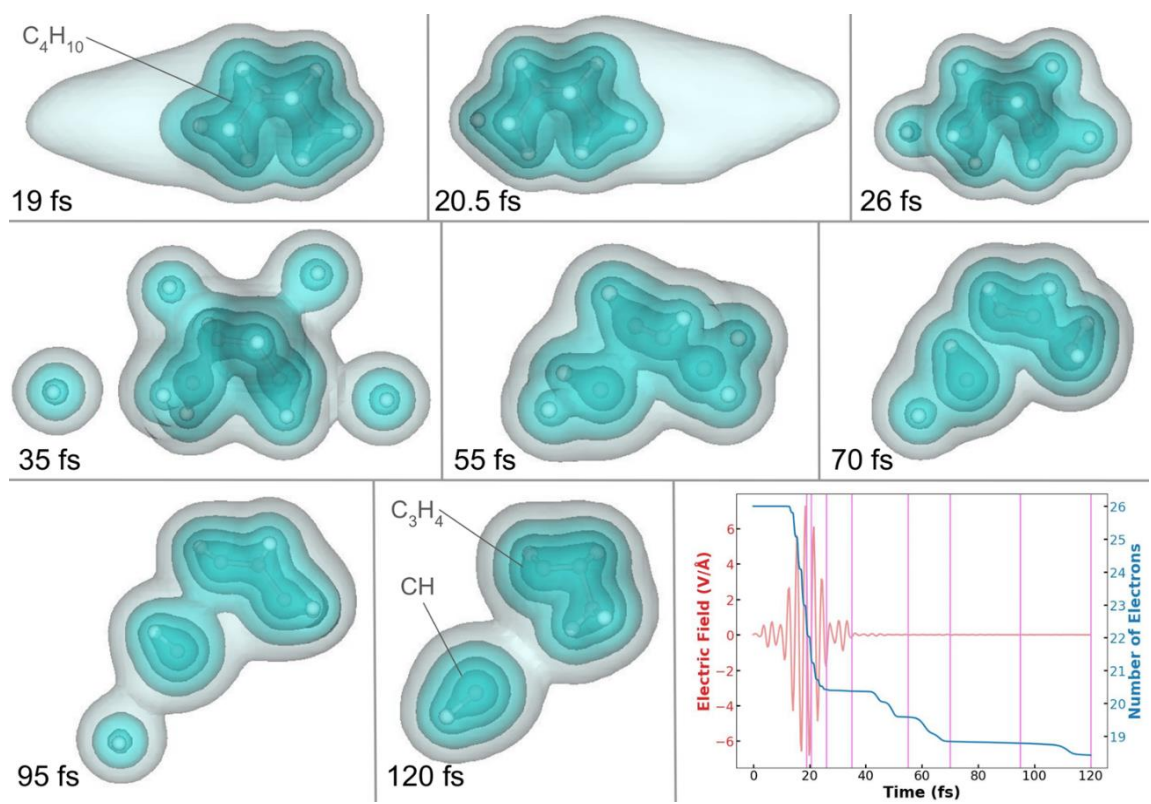

**Figure 8.** Coulomb explosion snapshots from the simulation resulting in the formation of CH and $C_3H_4$ molecules measured to have 4.30 and 14.12 valence electrons and $0.70^+$ and $1.88^+$ charges, respectively (the 0.5, 0.1, 0.01, and 0.001 density isosurfaces are shown)

The simulation results are presented with two additional snapshot series with the formation of one $C_2$ and two $CH_2$ fragments (Figure S17a) and $C_3H_2$ and hydrogen fragments (Figure S17b). The Figure 8 shows that rapid ionization starts when the laser strength reaches about 5 V/Å (16 fs). Electron density is pulled away from the nuclei and is removed by the complex absorbing potential. As the electric field changes direction with the carrier-envelope phase, the density is pulled back through the molecule, where it interacts with all the spin orbitals via the Hartree and exchange-correlation potentials. The molecule loses - 5 or 6 of its 26 valance electrons by the time the laser strength decreases below 5 V/Å (26 fs). After the laser pulse is too weak to ionize the system, the number of electrons remains constant up to 50 fs. Then the electron count starts to decrease because the ejected atoms leave the simulation box. The bonds stretch as the Coulomb forces between the nuclei are too great for the remaining electrons to stabilize. After about 100 fs, the system consists of slowly moving charged fragments.

The histogram represents the fragmentations of the $C_4H_{10}$ molecule starting from 88 different random velocities (Figure 9). Hydrogen is the most frequently (75%) produced fragment and has an average charge of $0.6^+$ at the end of the 120 fs simulation window. Experimentally, its presence in neutral form was detected with high intensity in the emission spectra (Figure 2a and Figure S7). Based on the simulations, the most heavily produced carbon–containing fragments are $CH_2$ radicals, but many other fragments, including $C_2H_2$, $C_2$, and CH, can be formed after Coulomb explosion mostly in a positively charged form (Figure 9). The average appearance frequency of CH molecules was 0.5% among all generated fragments in 88 simulation runs (Figure 9). The distribution of the charge states of each carbon–containing fragment after the Coulomb explosion are shown in Figure S18. The height of the columns shows the frequency of the fragment and the numbers above each column represent the frequency of the different products. Each of the columns are subdivided into bin sizes of 0.2 charge unit. The different colors represent different charge states.

The calculated remaining number of valence electrons of CH is 4.4, thus the average charge is +0.6. Therefore, CH molecules have a 40% chance to be neutral and 60% chance of being $1^+$ ion after 120 fs.

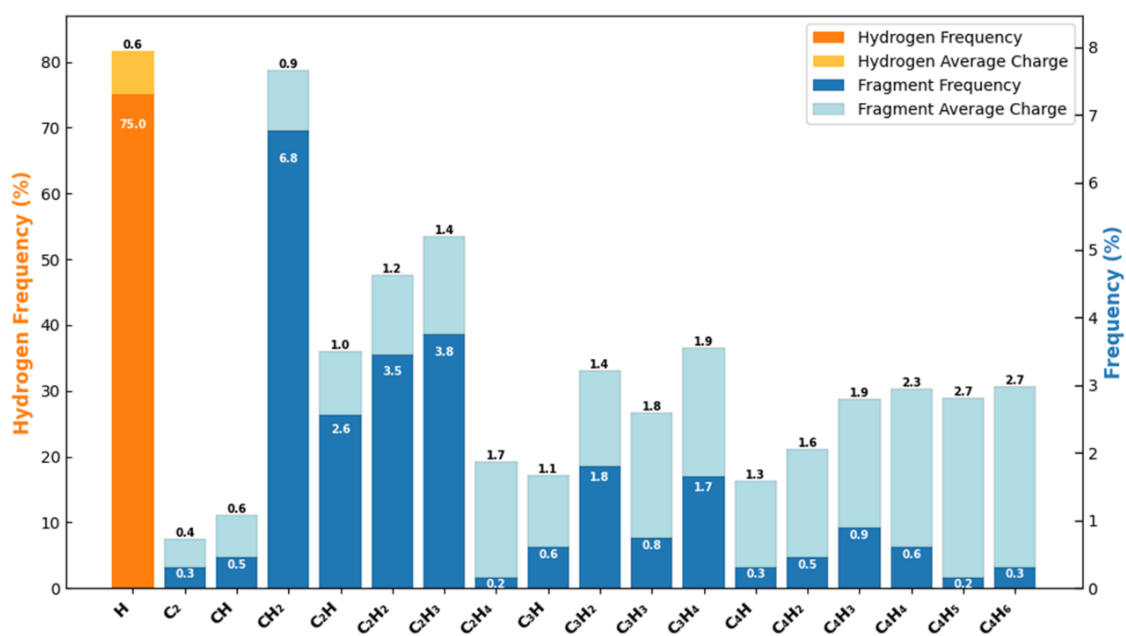

**Figure 9.** The average appearance frequency of each fragment. Blue columns are representing the frequency of carbon containing fragments and the orange column the hydrogen fragments. The lighter colored appended columns are showing the average charges

The CH(A) emission was detected in the first few nanoseconds in our experiments at a higher peak intensity than the observation threshold of the CH fragments in the simulations. This may be due to the fact that we collect fluorescent light not only from the focal point but also from a larger distance along the bright laser filament (Figure S6). The estimated peak intensity is decreasing to the applied peak intensity value in the simulations ($7 \times 10^{14}$ W/cm$^2$) in the distance of about 1.3 mm from the focal spot in the direction of the beam propagation. This results in a spectrum that involves the light from regions in which the actual peak intensity is significantly lower compared to the calculated value in the focal spot. However, a potential mechanism may exist for CH$^+$ ions to be formed and neutralized in the first few hundred picoseconds.

CONCLUSIONS

The CH(A), C$_2$ radicals and H atoms were efficiently generated from butane-argon plasma with the SYLOS laser (8 fs, 760 µJ/pulse, 1 kHz, 890 nm). The CH(A) spectrum was measured with high resolution with the step-scan FTVis measurement technique. The characteristic features of the CH(A) emission band were recognizable in the first 5 ns time point indicating the early formation of CH(A) radicals. The GDD value of the laser pulse between $-1000$ fs$^2$ and $+1000$ fs$^2$ had a significant influence on the intensity and composition of the observed spectra. The simulations of the argon emission GDD dependent profile at $2\times10^{16}$ W/cm$^2$ was in good agreement with the measured data. All investigated spectral components showed the highest intensity at GDD 0 fs$^2$ and decreased gradually in the studied GDD range in the sensitivity order of Ar$^+$ > C$_2$ > H-α ~ CH(A). We concluded that not only the value of the GDD is important but also its sign has an impact on the spectral composition. All CH(A) emission decay curves can be fitted with a monoexponential function when the spectra are argon emission background corrected. The initial vibrational temperatures at 75 ns were in the range of 3770 – 3850 K, and the rotational temperatures were measured between 2870 and 3135 K. The rotational temperatures decreased from ~3000 K to 2000–2400 K and the vibrational temperatures increased from ~3800 K to 4900 K only at 0 fs$^2$ GDD. It can be assumed that this increase is due to some secondary exothermic chemical reactions on the longer timescale. Our TDDFT simulations predict that many fragments (e.g. $CH_2$, $C_2H$, CH) can be formed from butane in 120 fs. CH molecules with $0.6^+$ average charge were predicted in these calculations at $7\times10^{14}$ W/cm$^2$ peak intensity. The CH(A) emission measured experimentally between the $-1000$ and $+1000$ fs$^2$ GDD range confirmed that the direct formation of neutral CH is possible at such high applied peak intensities ($1.8\times10^{14}$–$1.2\times10^{16}$ W/cm$^2$). The 1.3-1.6 ns rise times of the CH(A) emission further supports the conclusion that these radicals are mostly formed directly in the Coulomb explosion process or in the subsequent few hundred picosecond time range when the electron density is still high in the plasma.

The application of the GDD tunable femtosecond laser pulses may open new possibilities for the light-driven control of chemical reactions in gas phase.

ACKNOWLEDGEMENTS

The ELI ALPS project (GINOP-2.3.6-15-2015-00001) is supported by the European Union and co-financed by the European Regional Development Fund. This work was supported by the National Science Foundation (NSF) under Grants No. NSF IRES 2245029.

TABLE OF CONTENTS

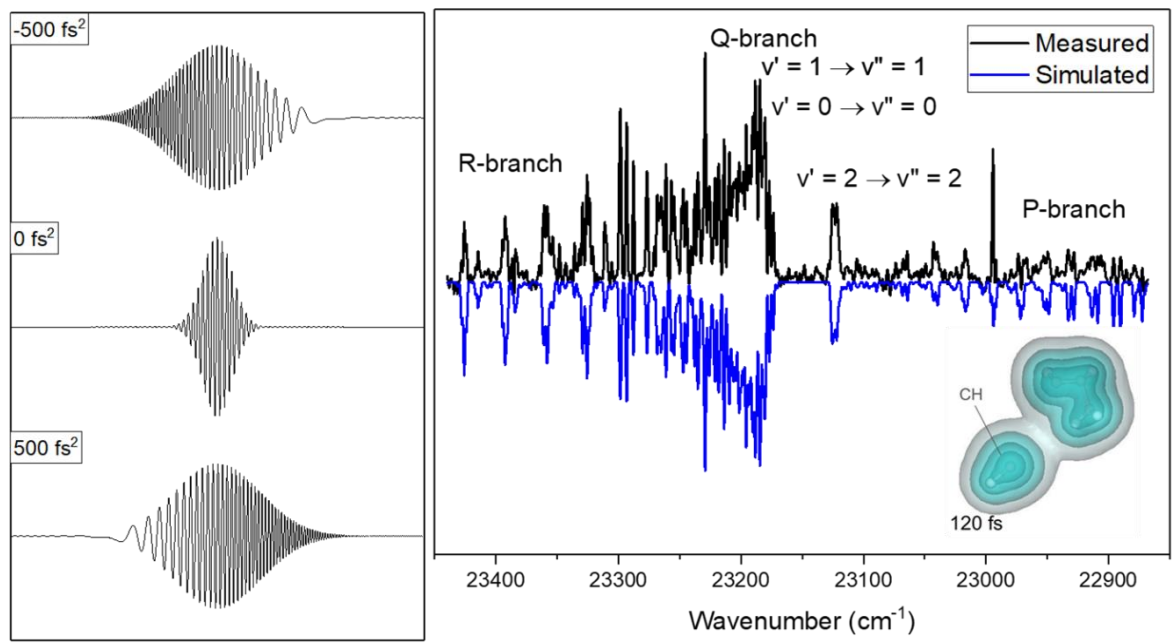

# Supporting Information

*Karoly Mogyorosi[1], Balint Toth[1], Krisztina Sarosi[1,2], Barnabas Gilicze[1], Janos Csontos[1], Tamas Somoskoi[1], Szabolcs Toth[1], Prabhash Prasannan Geetha[1], Laszlo Toth[1], Samuel S. Taylor[3], Nicholas Skoufis[3], Liam Barron[3], Kalman Varga[3], Cody Covington[4], Viktor Chikan[1,5,6]*

[1]ELI ALPS, ELI-HU Non-Profit Ltd, Wolfgang Sandner u. 3, H-6728 Szeged, Hungary

[2]Department of Optics and Quantum Electronics, University of Szeged, Dóm tér 9, H-6720 Szeged, Hungary

[3]Department of Physics and Astronomy, Vanderbilt University, Nashville, Tennessee, 37235, USA

[4]Department of Chemistry, Austin Peay State University, Clarksville, USA

[5]Department of Chemistry, Kansas State University, Manhattan, Kansas 66506-0401, United States

[6]ASML, 17082 Thornmint Ct, San Diego, CA 92, United States

(a)

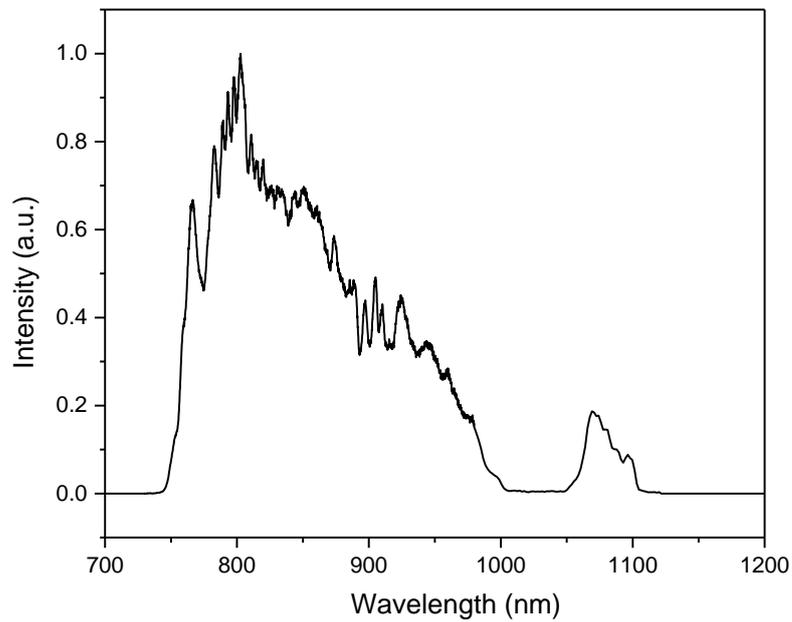

(b)

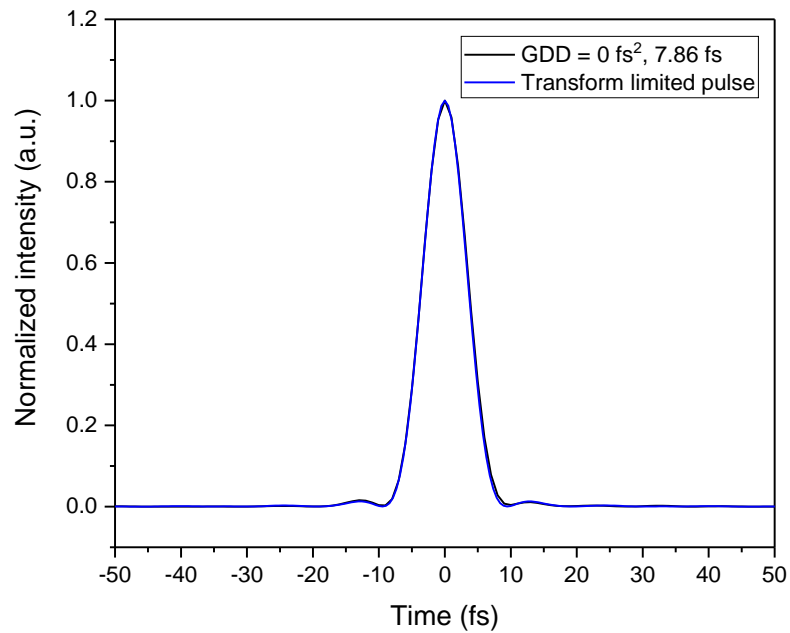

**Figure S1.** (a) Output spectrum of SYLOS2 laser source (b) retrieved temporal profile from pulse duration measurements using the chirp scan technique (FWHM = 7.86 fs) and its transform limited temporal profile (FWHM = 7.86 fs)

The temporal distribution of the chirped pulses calculated from the retrieved complex spectrum is shown in Figure S2.

(a)

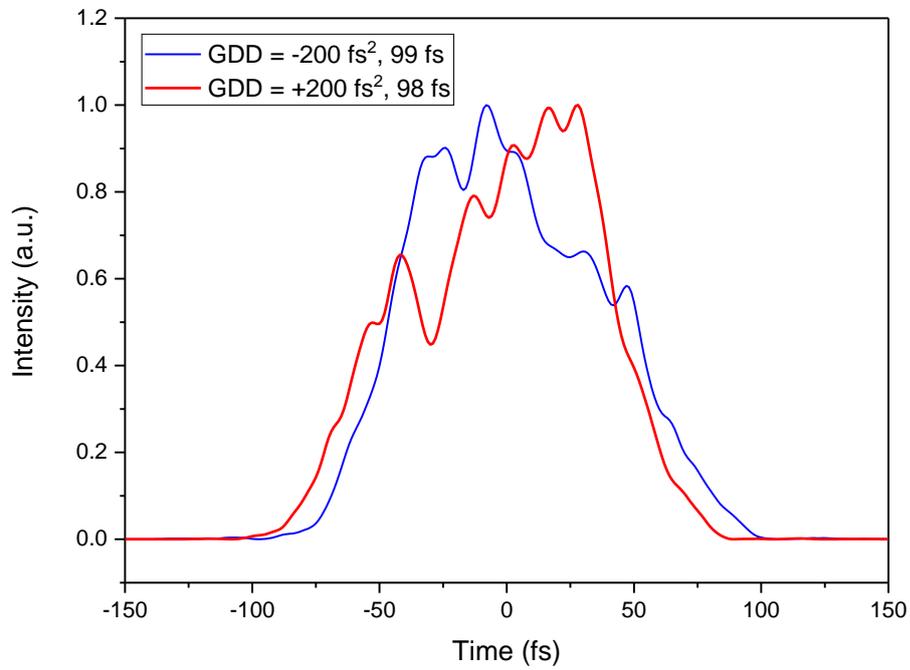

(b)

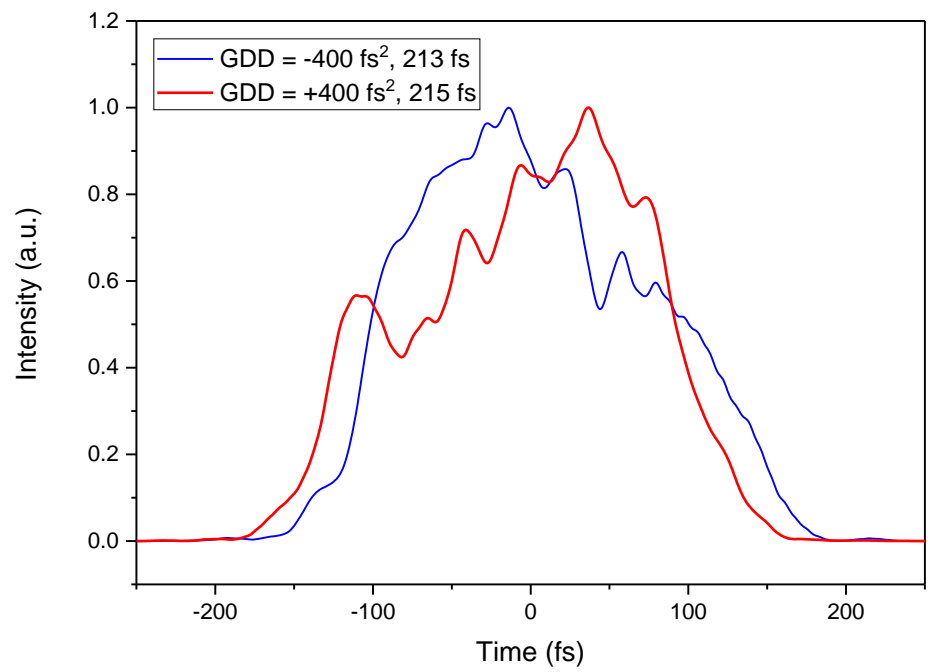

(c)

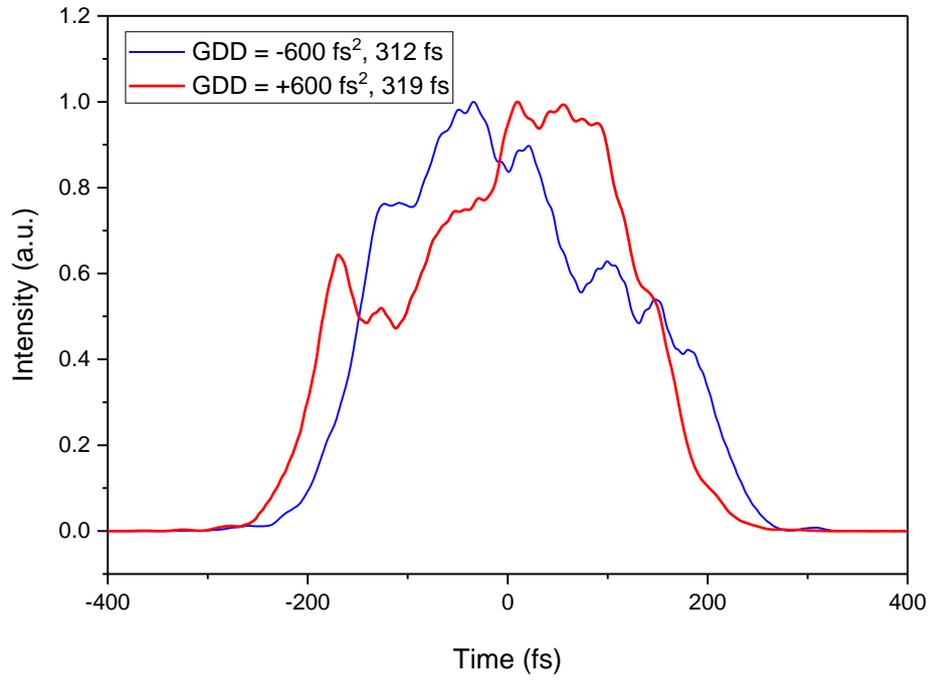

(d)

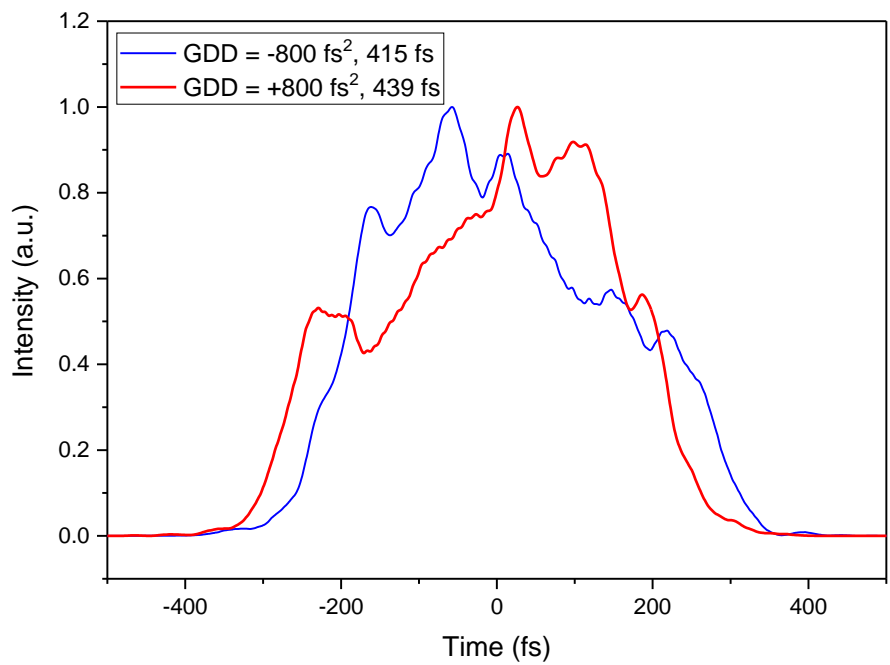

(e)

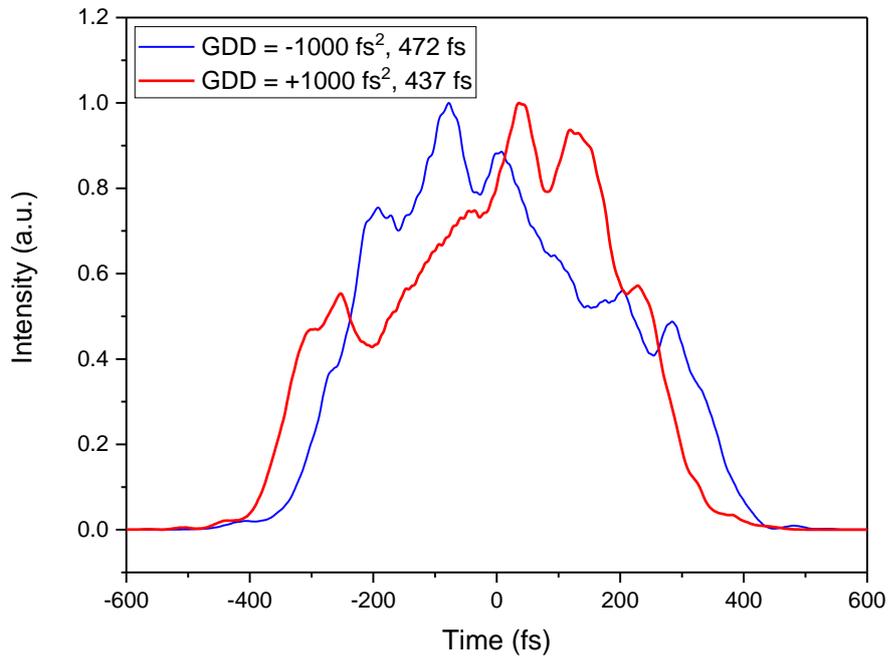

**Figure S2.** Pulse duration determination at different GDD values: a) -200 and +200 fs², b) -400 and +400 fs², c) -600 and +600 fs², d, -800 and +800 fs² and e, -1000 and +1000 fs²

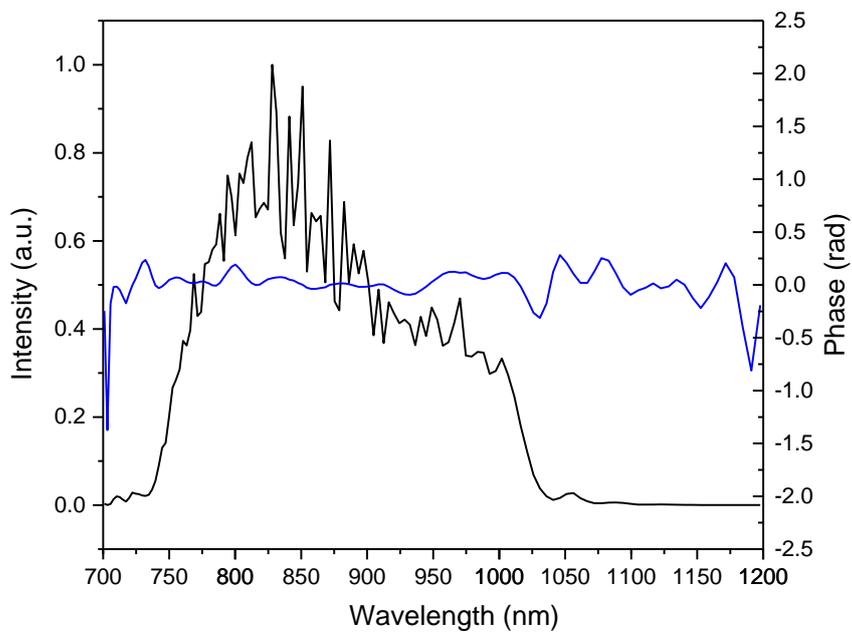

**Figure S3.** Retrieved output spectrum and spectral phase of SYLOS2 from the chirp scan measurements

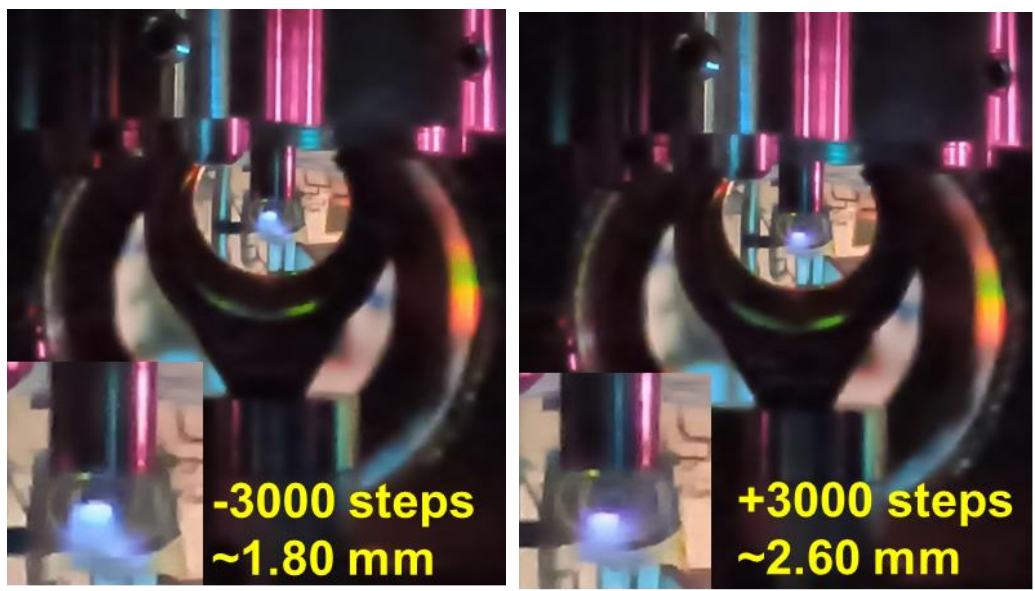

**Figure S4.** Plasma images recorded to determine the piezo valve–plasma distances at two different picomotor settings by tilting the focusing spherical mirror

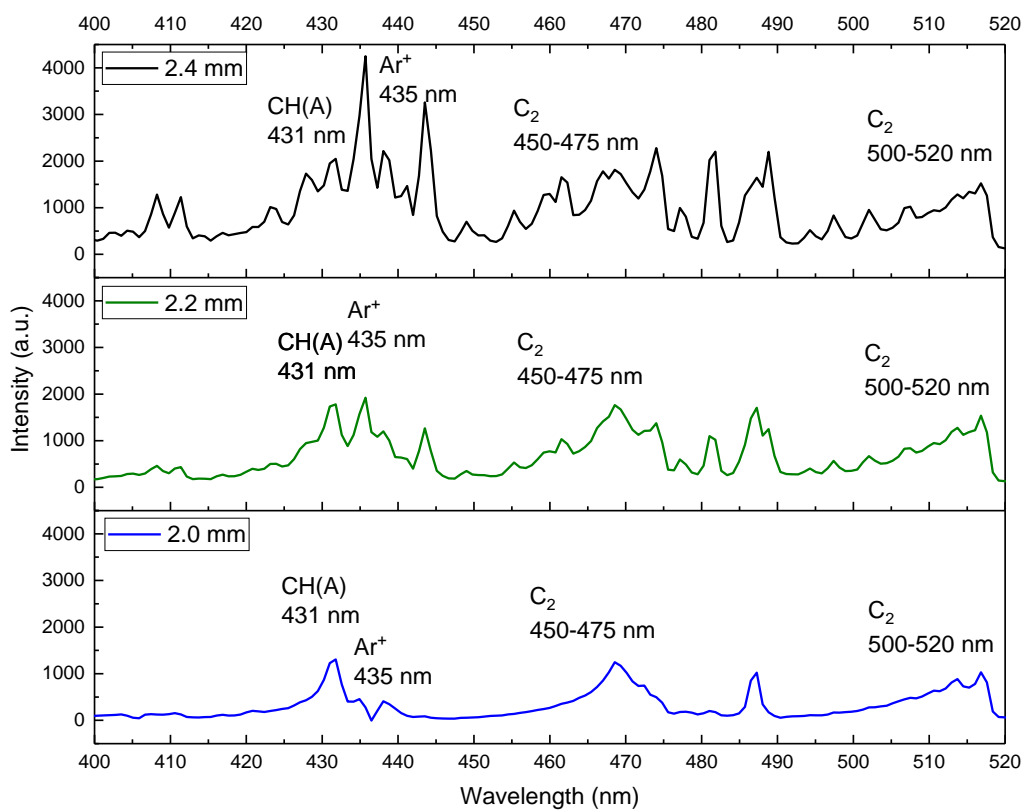

**Figure S5.** Overview emission spectra measured from the brightest butane-argon plasma at three different piezo valve–focal point distances

At a given GDD value, we measured the average overview emission spectrum with the QEPro spectrometer (100 ms integration time, average of 10 spectra).

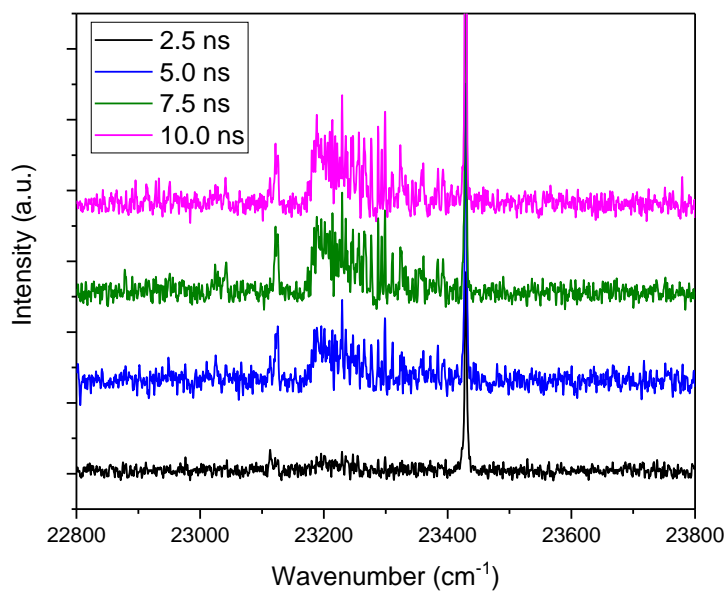

**Figure S6.** CH(A) emission spectra from methane-helium plasma at the early time points between 2.5 and 10.0 ns (spectra shifted for better visibility) measured at GDD 0 fs$^2$

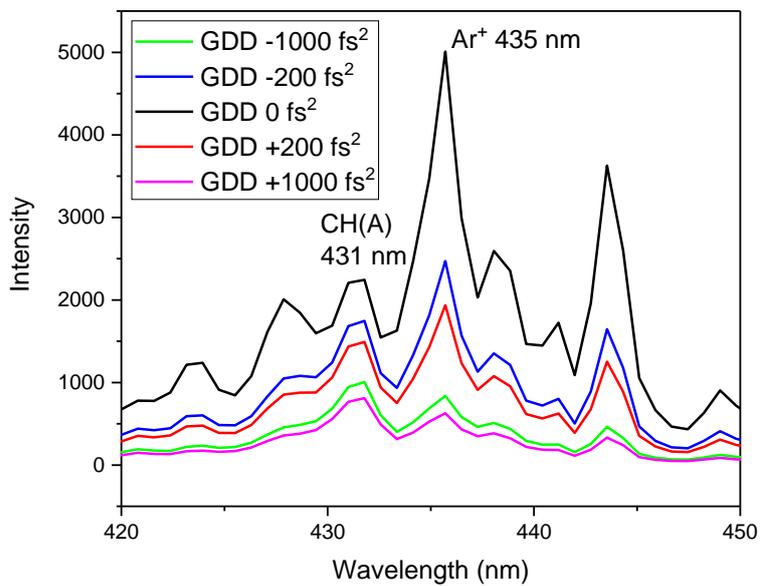

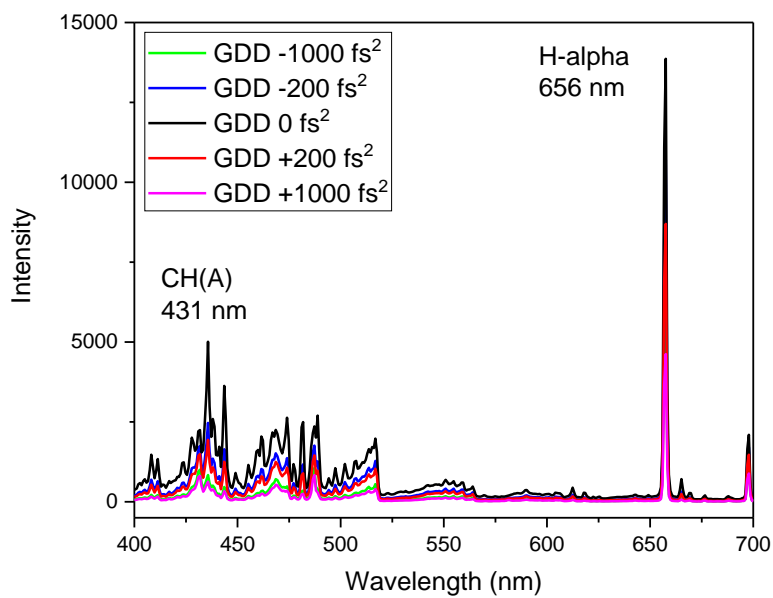

**Figure S7.** Overview of emission spectra from butane-argon plasma at different GDD values a, in the 420-450 nm and b, in the 400-700 nm wavelength range

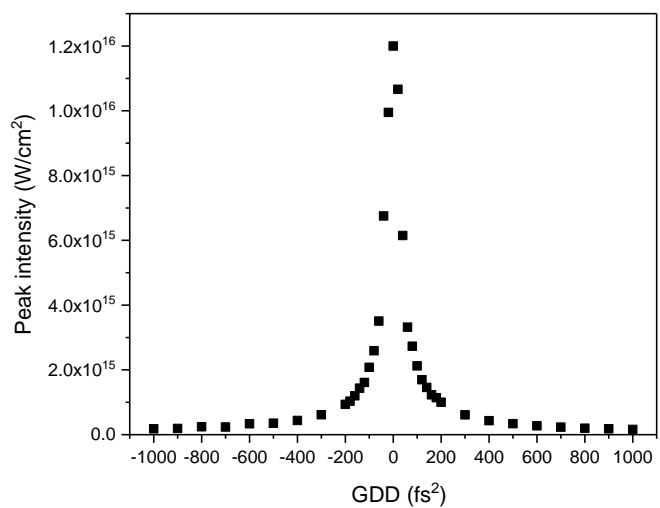

**Figure S8.** The GDD dependent calculated peak intensity

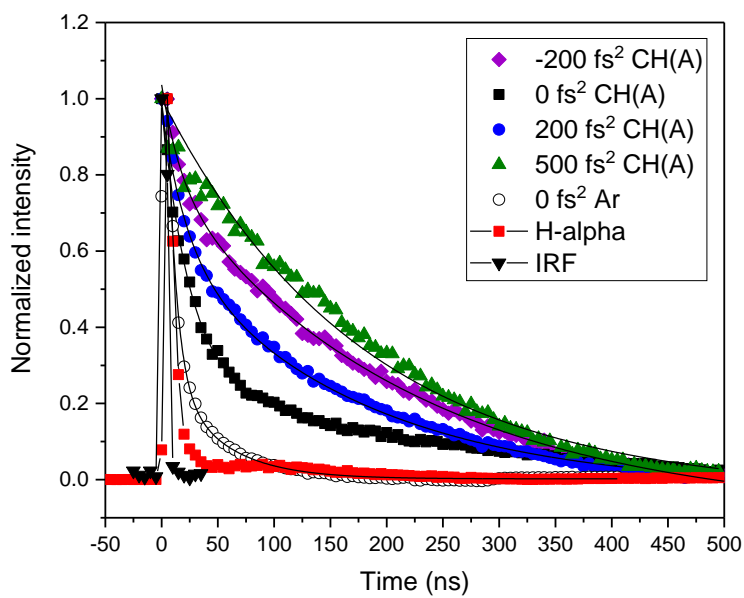

**Figure S9.** The GDD dependent exponential decay curves of the CH(A) emission integrated in the 23,175–23,310 cm$^{-1}$ spectral range at –200 fs$^2$, 0 fs$^2$, 200 fs$^2$ and 500 fs$^2$ GDD without argon background correction and the argon emission decay curve at 22,994–23,002 cm$^{-1}$, the H-alpha emission decay curve at 15,240-15,250 cm$^{-1}$ and the Instrument Response Function (IRF)

(a)

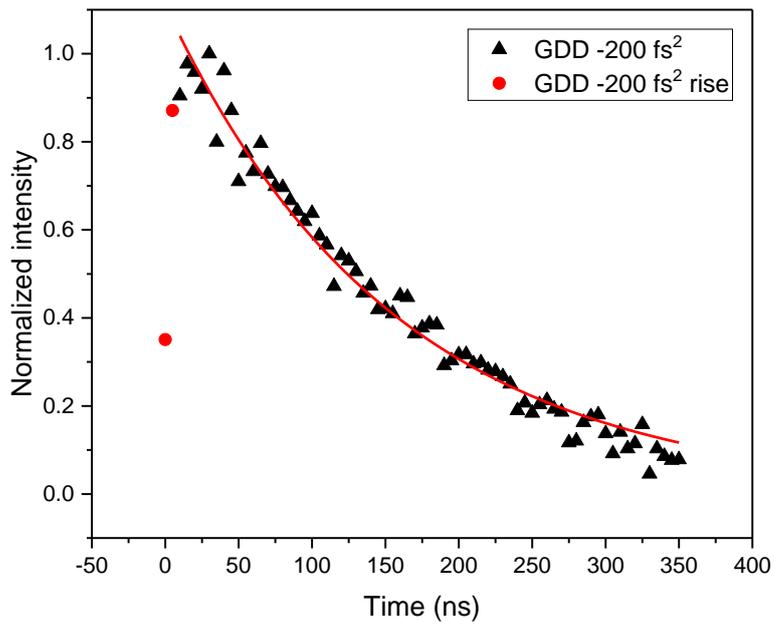

(b)

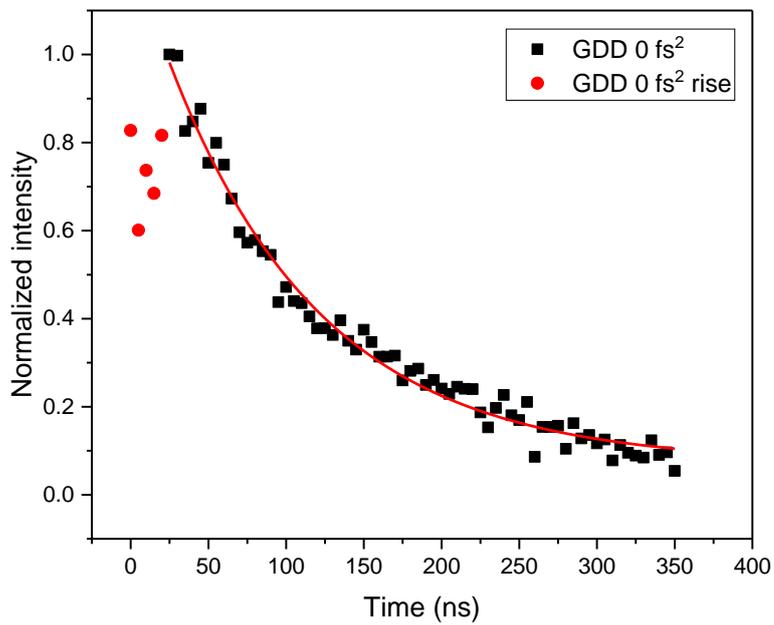

(c)

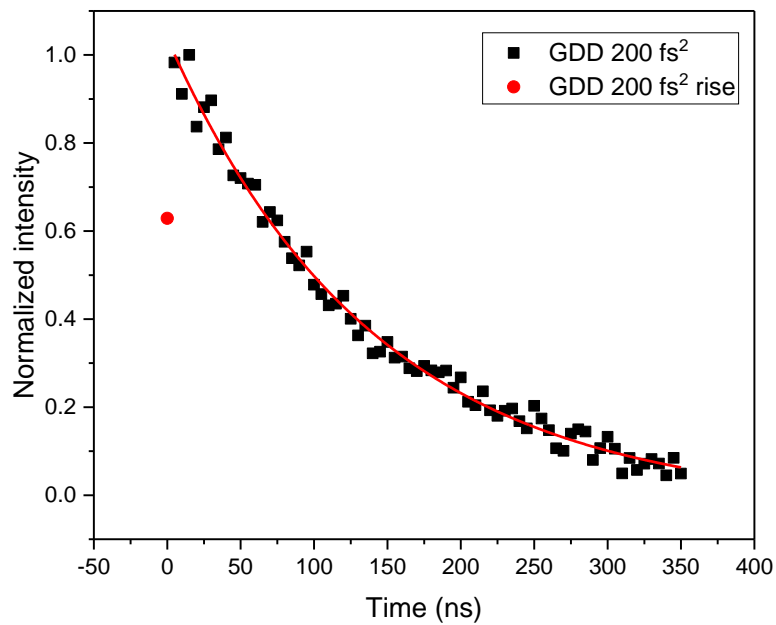

(d)

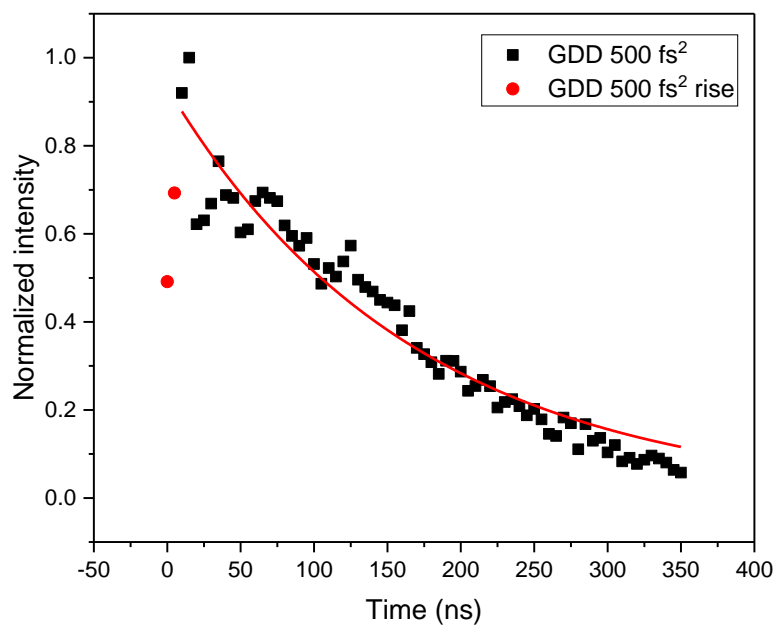

**Figure S10.** The CH(A) emission decay curves fitted with monoexponential functions at (a) –200 fs$^2$, (b) 0 fs$^2$, (c) 200 fs$^2$ and (d) 500 fs$^2$ GDD

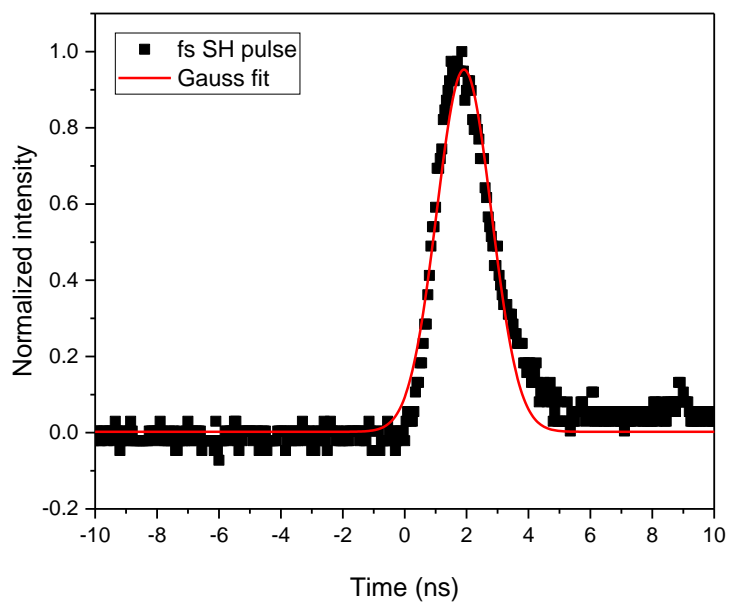

**Figure S11.** The Instrument Response Function determination with the fast PMT (rise time: 0.5 ns), the Ortec pre-amplifier (rise time 300 ps) and the Tektronix oscilloscope (MSO 71254C, 12.5 GHz) applying the ultrashort Second Harmonic (SH) laser pulses of the HR1 alignment laser (7 fs)

a,

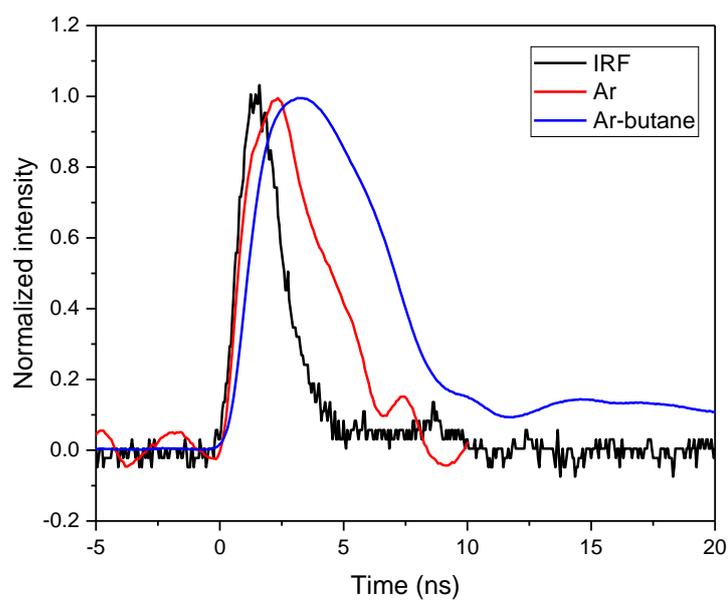

b,

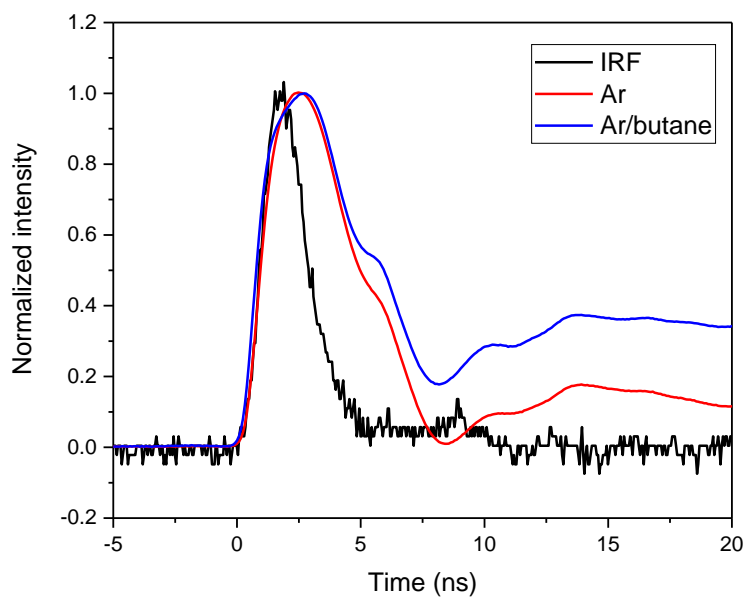

**Figure S12.** The normalized CH(A) emission signal evolution from butane-argon plasma using the HR1 alignment laser (1030 nm, 1 kHz, 7 fs, ~9×10$^{15}$ W/cm$^2$) and FB430-10 filter a, pulsed measurement mode (3 bar) and b, static measurement mode (1 mbar)

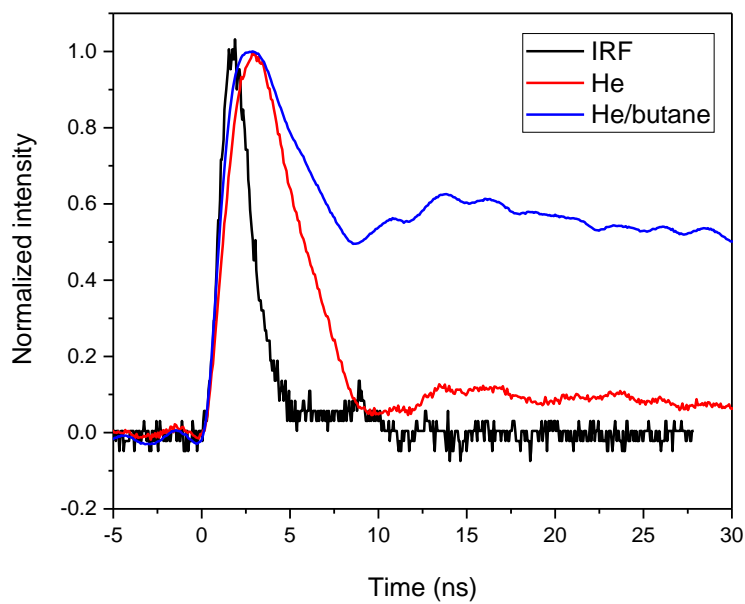

**Figure S13.** The normalized CH(A) emission signal evolution from helium and butane-helium plasma using the HR1 alignment laser and FB430-10 filter in static measurement mode (3 mbar)

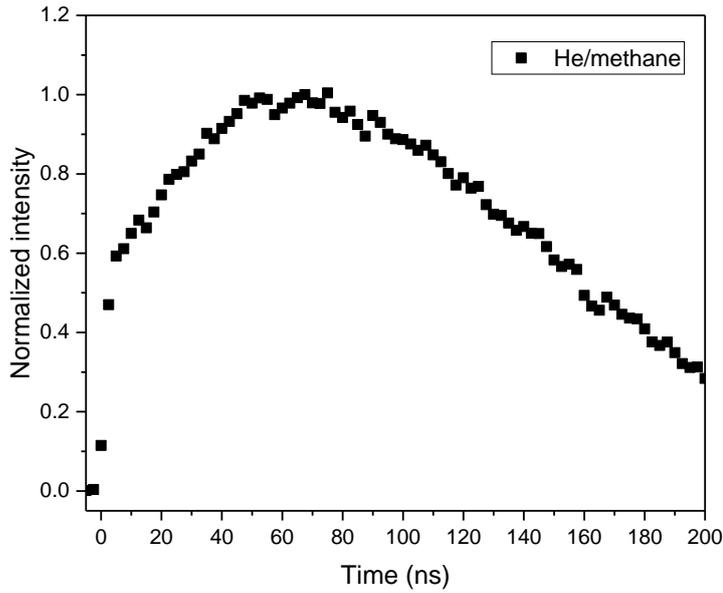

**Figure S14.** CH(A) emission kinetics from methane-helium plasma measured at GDD 0 fs$^2$

**Collision time calculations**

The mean free path of a single atom gas can be calculated as[48]:

$$\lambda = \frac{kT}{\sqrt{2}\sigma p}$$

eq 1

where k is the Boltzmann constant, T is the temperature, p is the pressure, σ is the collision cross section calculated from the van der Waals diameter of argon. Considering from the ideal gas law that

$$\frac{N}{V} = \frac{p}{kT}$$

eq 2

the mean free path can be written as

$$\lambda = \frac{V}{\sqrt{2}\sigma N}$$

eq 3

where

$$\sigma = \pi d^2$$

For argon, we consider the p = 30 mbar (at 2.2 mm distance from the nozzle) and the temperature is taken as an average of the initial vibrational and rotational temperatures that were calculated from the spectrum simulations, such as T = 3500 K. It should be noted that the translational temperature could be different from the vibrational and rotational temperatures as the molecules are not in equilibrium. The van der Waals radius of argon is 188 pm. The mean velocity of the atoms can be calculated as:

$$v = \sqrt{8kT/\pi m}$$

eq 5

where m is the mass of the atom. With the above parameters, the v = 1378.6 m/s and the $\lambda$ = $2.56 \times 10^{-5}$ m. The mean collision time is 18.6 ns for the argon atoms with the above mentioned parameters.

We can also estimate the ion-ion collision time based on the following equation[49]:

$$v_{ii} = \frac{1}{3\pi^{1/2}} n_i \left(\frac{q_1 q_2}{4\pi\varepsilon_0}\right)^2 \frac{4\pi}{m_i^{1/2} T_i^{3/2}} ln\Lambda$$

eq 6

where $n_i$ is the ion density, $q_1$ and $q_2$ are the charges of the ions ($1.60 \times 10^{-19}$ C), $m_i$ is the mass of the ion and $T_i$ is the temperature of the ion and $ln\Lambda$ = 15. Based on the simulation results, we calculated the velocity of the $CH_2^+$ ion to be 6250 m/s and the $T_i$ = 25800 K from eq5. The density of pure argon in the 2.2 mm distance from the nozzle was estimated to be $n_{Ar}$ = $5.16 \times 10^{22}$ $1/m^3$ and considering the 3 v/v% butane content, the maximum $n_i$ = $6.19 \times 10^{21}$ $1/m^3$. The $m_i$ = $2.33 \times 10^{-26}$ kg and the $v_{ii}$ = $3.61 \times 10^8$ 1/s. The collision time of the $CH_2^+/CH_2^+$ collisions is $2.77 \times 10^{-9}$ s = 2.77 ns.

(a)

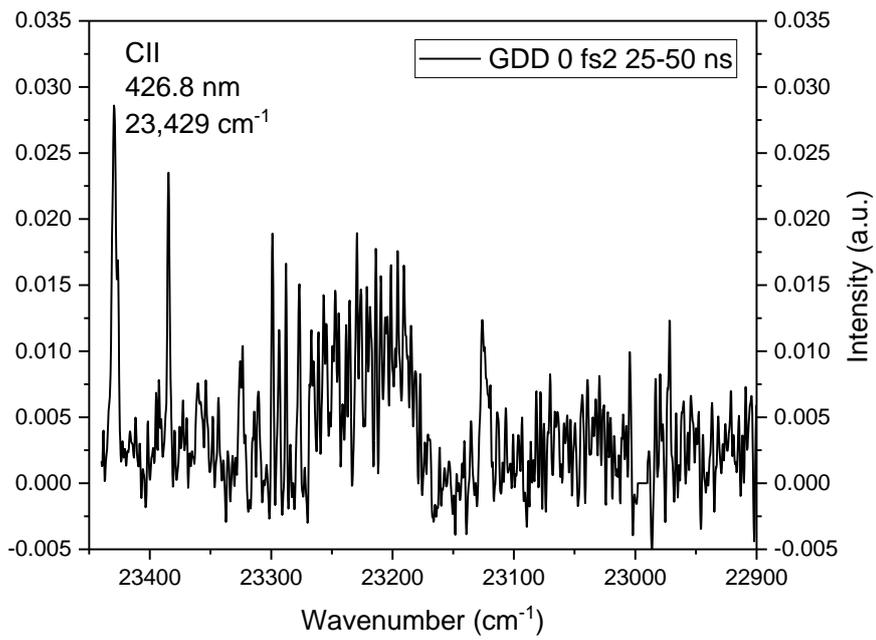

(b)

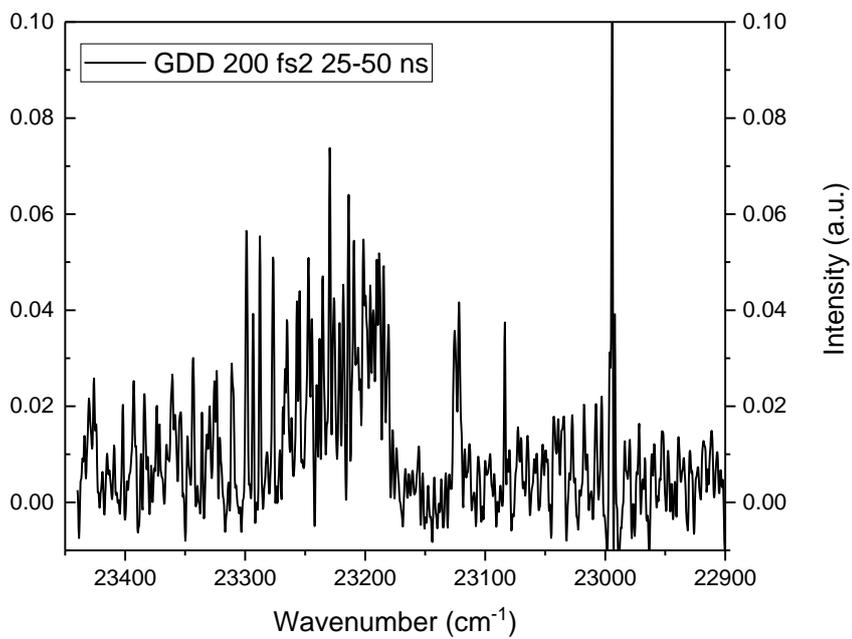

(c)

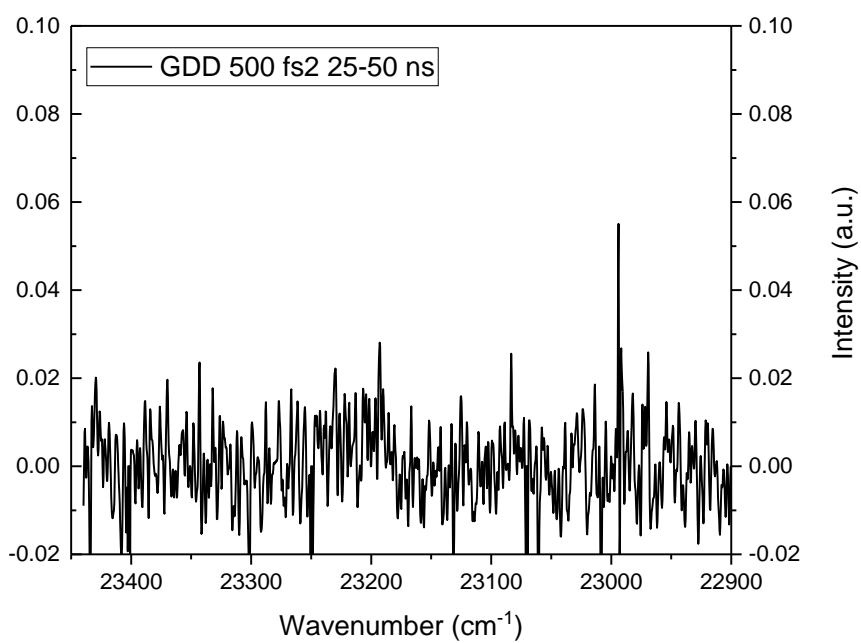

**Figure S15.** High resolution spectra of early time points (25–50 ns) at (a) 0 fs$^2$, (b) +200 fs$^2$ and (c) +500 fs$^2$ GDD

## TDDFT SIMULATIONS

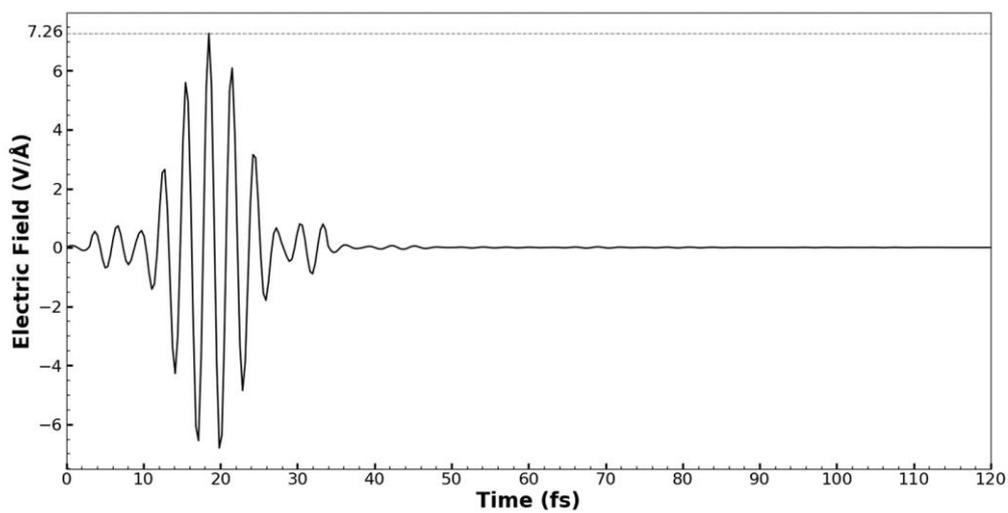

**Figure S16.** The pulse used in the 120 fs butane Coulomb explosion simulations (GDD 0 fs$^2$)

(a)

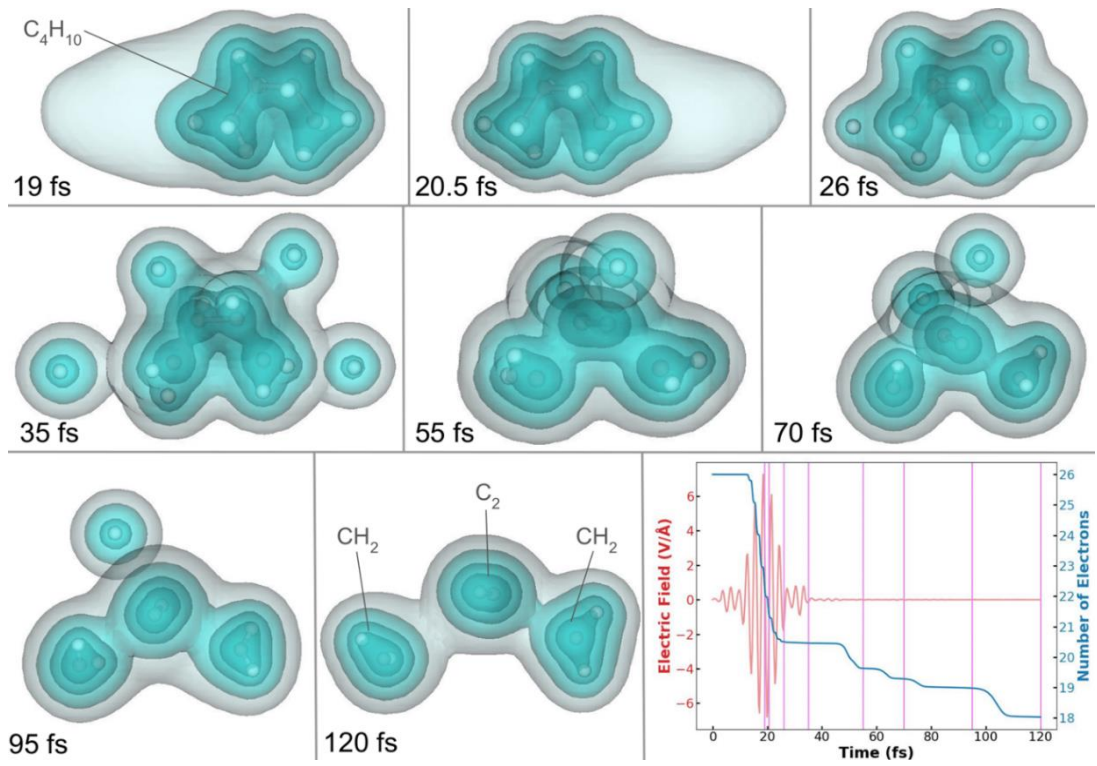

(b)

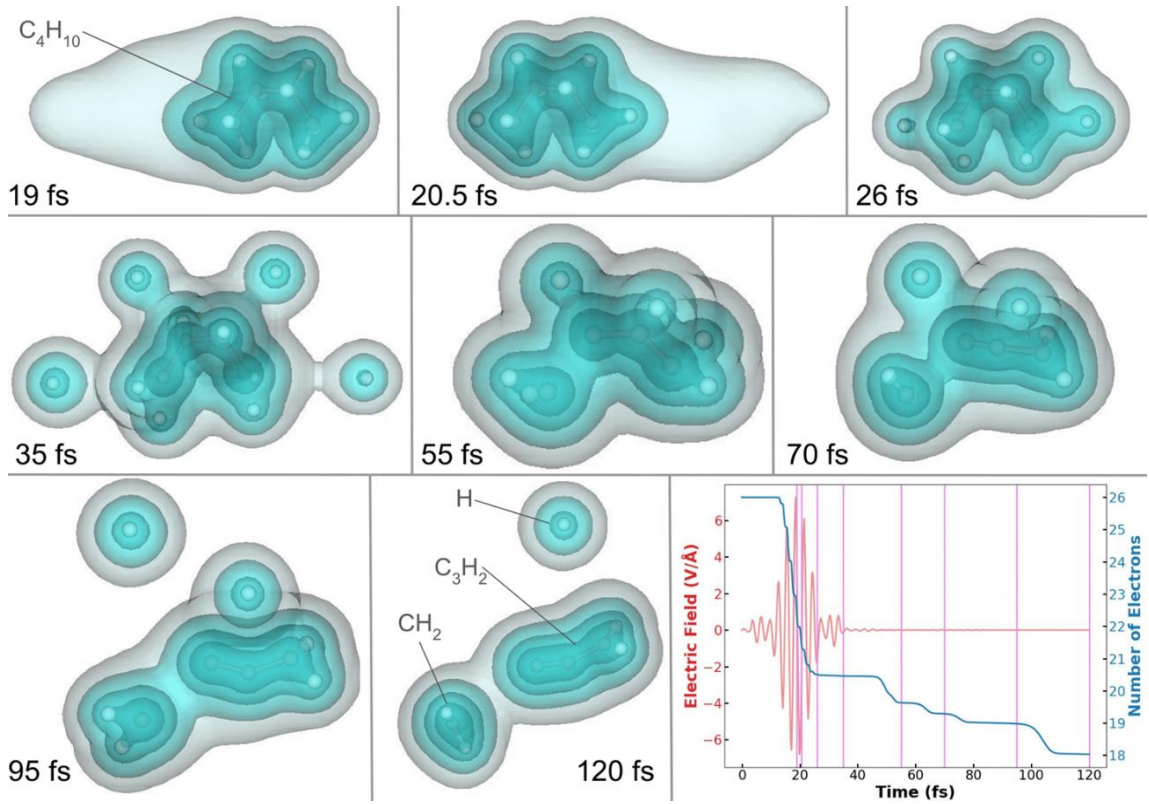

**Figure S17.** Coulomb explosion snapshots from the simulations resulting in the formation of: (a) $C_2$, and two $CH_2$ molecules measured to have 7.53, 5.23, and 5.26 valence electrons and $0.47^+$, $0.77^+$ and $0.74^+$ charges, respectively and (b) $CH_2$, $C_3H_2$ molecules and H atoms measured to have 5.03, 12.63, and 0.48 valence electrons and $0.97^+$, $1.37^+$ and $0.52^+$ charges, respectively. The 0.5, 0.1, 0.01, and 0.001 density isosurfaces are shown.

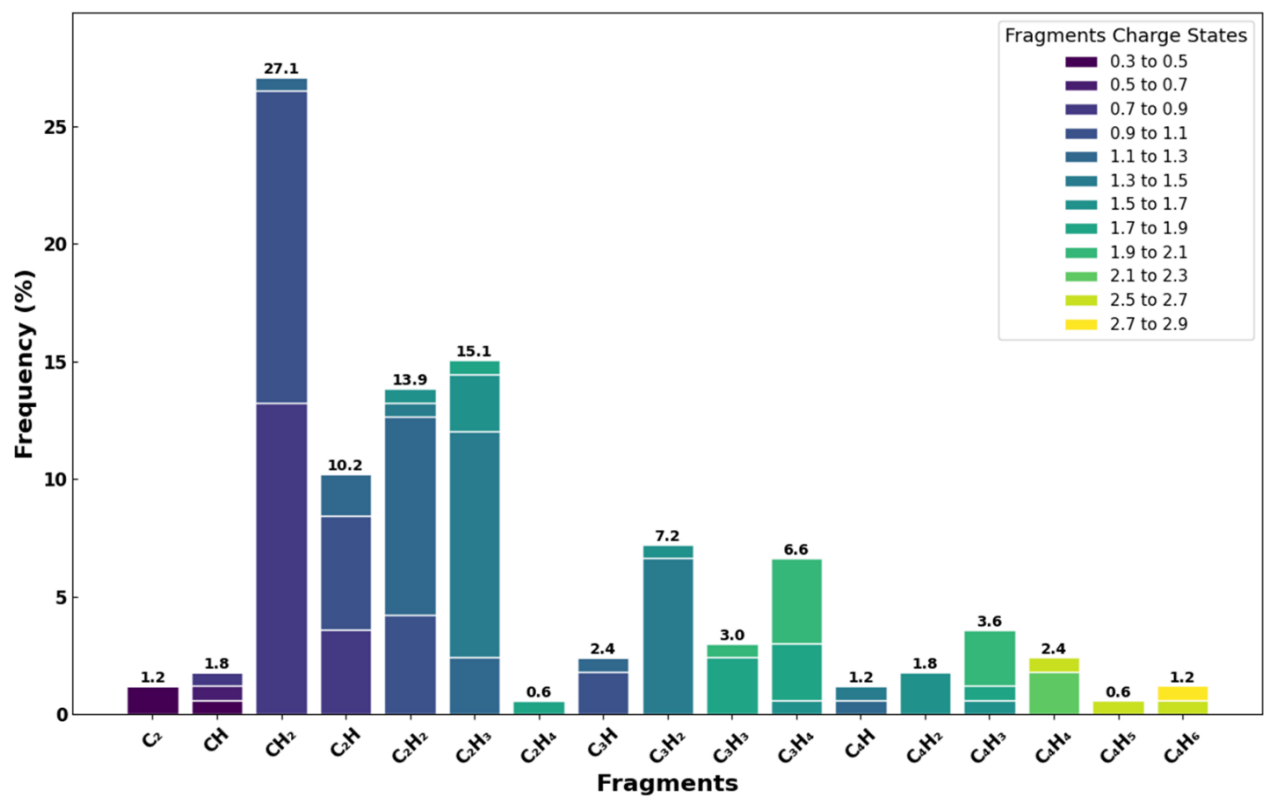

**Figure S18.** The distribution of each fragment's charge states after the Coulomb explosion